\documentclass[]{eptcs}
\providecommand{\event}{WWV'12} 
\usepackage{breakurl}             
\usepackage{makeidx}  
\usepackage{amsmath}
\usepackage{amsthm}
\usepackage{amssymb}
\usepackage[small,nohug,heads=littlevee]{diagrams}
\diagramstyle[labelstyle=\scriptstyle]
\usepackage[dvips]{color}
\usepackage{tikz}
\usetikzlibrary{positioning,shapes,shadows,arrows}

\newtheorem{myprop}{Property}
\newtheorem{mydef}{Definition}[section]

\title{Adding Sessions to BPEL}
\author{Jonathan Michaux
\institute{T\'{e}l\'{e}com ParisTech\\ Paris, France}
\email{michaux@telecom-paristech.fr}
\and
Elie Najm
\institute{T\'{e}l\'{e}com ParisTech\\ Paris, France}
\email{najm@telecom-paristech.fr}
\and
Alessandro Fantechi
\institute{Universit\`{a} degli Studi di Firenze\\ Florence, Italy}
\email{fantechi@dsi.unifi.it}
}
\def\titlerunning{Adding Sessions to BPEL}
\def\authorrunning{J. Michaux, E. Najm \& A. Fantechi}

\begin{document}
\maketitle


\newcommand{\MM}[1]{\ensuremath{#1}}
\newcommand{\REGLE}[3]{\box{
                                           \MM{\dfrac{\begin{array}{l} #1 \end{array}}
                               {\begin{array}{l} #2 \end{array}}}
                               }
                               \vspace{1mm}
                            \begin{left} \hbox{#3} \end{left}
                     }
\newcommand{\regle}[2]{\MM{\frac{\begin{array}{c} #1 \end{array}}
                               {\begin{array}{l} #2 \end{array}}}}
\newcommand{\formatreg}[1]{\begin{center} \fbox{#1} \end{center} \vspace{5mm}}
\newcommand{\Fait}[1]{\MM{\ \ \xrightarrow{\ #1\ }}\ \ }
\newcommand{\fait}[1]{\MM{\xrightarrow{\ #1\ }}}
\newcommand{\faitpas}[1]{\MM{\ \ \ \ {\scriptsize \not}\!\!\!\!\!\!\!\xrightarrow{\ #1\ }}}
\newcommand{\RL}[1]{\,\frac{}{\,^{{#1}}\,}\,}
\newcommand{\RS}[1]{\,\frac{}{^{{#1}}}\,}

\newcommand{\sqplus}{\ {\scriptstyle {\widehat{+}}}\ }
\newcommand{\smplus}{\ {\scriptstyle {\dot{+}}}\ }

\newcommand{\intrupt}{\,{\scriptstyle\leadsto}\,}

\newcommand{\rget}{\frac{}{\text{\ttfamily\ Get\ }}}
\newcommand{\rset}{\frac{}{\text{\ttfamily\ Set\ }}}
\newcommand{\rgvt}{\MM{ \frac{}{\text{\ttfamily\  Get\_Event\ }} }}
\newcommand{\rsvt}{\frac{}{\text{\ttfamily\ Send\_Event\ }}}
\newcommand{\mgvt}{\frac{}{\text{\ttfamily\ \ Get\_Event\ }}}
\newcommand{\msvt}{\frac{}{{\text{\ttfamily\ \tiny Send\_Event\ }}}}

\newcommand{\rx}{\frac{}{\large \ x \ }}
\newcommand{\smcirc}{{\scriptscriptstyle \circ}}
\newcommand{\smodot}{{\scriptscriptstyle \odot}}
\newcommand{\smoplus}{\,{\scriptstyle \oplus}\,}
\newcommand{\smbar}{{\scriptscriptstyle \|}}
\newcommand{\smla}{ {\scriptscriptstyle \langle } }
\newcommand{\smra}{ {\scriptscriptstyle \rangle } }
\newcommand{\sbullet}{\stackrel{}{\smla \:\!\! \smra}}

\newcommand{\mynil}{\text{{\footnotesize \bf \sffamily{nil}}}}

\newcommand{\pardec}{ {\scriptstyle {\bf {P \hspace{-0,7mm}A \hspace{-0,2mm}R}}} \!\; }
\newcommand{\seqdec}{ {\scriptstyle {\bf {S \hspace{-0,3mm}E \hspace{-0,2mm}Q}}} \!\; }
\newcommand{\unfdec}{ {\scriptstyle {\bf {U \hspace{-0,3mm}N \hspace{-0,2mm}F}}} \!\; }
\newcommand{\sesdec}{ {\scriptstyle {\bf {S \hspace{-0,3mm} E \hspace{-0,2mm} S}}} \!\; }
\newcommand{\invdec}{ {\scriptstyle {\bf {I \hspace{-0,2mm} N \hspace{-0,2mm} V}}} \!\; }
\newcommand{\repdec}{ {\scriptstyle {\bf {R \hspace{-0,15mm} E \hspace{-0,2mm} P}}} \!\; }
\newcommand{\recdec}{ {\scriptstyle {\bf {R \hspace{-0,15mm} E \hspace{-0,2mm} C}}} \!\; }
\newcommand{\picdec}{ {\scriptstyle {\bf {P \hspace{-0,2mm} I \hspace{-0,25mm} C}}} \!\; }
\newcommand{\boxdec}{ {\scriptstyle {\bf {B \hspace{-0,2mm} O \hspace{-0,25mm} X}}} \!\; }
\newcommand{\flodec}{ {\scriptstyle {\bf {F \hspace{-0,3mm}L \hspace{-0,2mm}O}}} \!\; }
\newcommand{\nildec}{ {\scriptstyle {\bf {N \hspace{-0,2mm} I \hspace{-0,25mm} L}}} \!\; }

\newcommand{\anydec}{ {\scriptstyle {\bf {*}}} \!\; }

\newcommand{\newpara}[1]{\noindent {\bf \small #1}}
\newcommand{\bpel}{\text{{\scriptsize BPEL}}}
\newcommand{\seb}{\text{{\small SeB}}}
\newcommand{\wsdl}{\text{{\scriptsize WSDL}}}
\newcommand{\sos}{\text{{\scriptsize SOS}}}
\newcommand{\swp}
{{{{\text{\bf \scriptsize \textit{S\hspace{-0.1mm}w\hspace{-0.1mm}P}}}}}}
\newcommand{\lb}{(\!\!(}
\newcommand{\rb}{)\!\!)}

\begin{abstract}

By considering an essential subset of the \bpel \hspace{0mm} orchestration language, we define \seb, a session based style of this subset. We discuss the formal semantics of \seb \hspace{0mm} and we present its main properties. We use a new approach to address the formal semantics, based on a translation into so-called control graphs. Our semantics handles control links and addresses the static semantics that prescribes the valid usage of variables. We also provide the semantics of collections of networked services.

Relying on these semantics, we define precisely what is meant by interaction safety, paving the way to the formal analysis of safe interactions between \bpel \hspace{0mm} services.

\end{abstract}
%

%

%

%
%
%

%

\section{Introduction}

In service-oriented computing, services are exposed over a network via well defined interfaces and specific communication protocols. The design of software as an orchestration of services is an active topic today. A service orchestration is a local view of a structured set of interactions with remote services.

In this context, our endeavour is to guarantee that services interact safely. The elementary construct in a Web service orchestration is a message exchange between two partner services. The message specifies the name of the operation to be invoked and bears arguments as its payload. An interaction can be long-lasting because multiple messages of different types can be exchanged in both directions before a service is delivered.

The set of interactions supported by a service defines its behavior. We argue that the high levels of concurrency and complex behavior found in orchestrations make them susceptible to programming errors. Widely adopted standards such as the Web Service Description Language (\wsdl \hspace{0mm}) \cite{wsdl2} provide support for syntactical compatibility analysis by defining message types in a standard way. However, \wsdl \hspace{0mm} defines one-way or request-response exchange patterns and does not support the definition of more complex behavior. Relevant behavioral information is exchanged between participants in human-readable forms, if at all. Automated verification of behavioral compatibility is impossible in such cases.

The present paper is a first step towards addressing the problem of behavioral compatibility of Web services. To that end, we follow the promising session based approach. Indeed, the session paradigm is now an active area of research with potential to improve the quality and correctness of software. We chose to adapt and sessionize a significant subset of the industry standard orchestration language \bpel \hspace{0mm} \cite{std/ws-bpel2}. We call the resulting language \seb \hspace{0mm} (Sessionized \bpel). It supports the same basic constructs as \bpel, but being a proof of concept, it does not include the non basic \bpel \hspace{0mm} constructs such as exception handling. These differences are explained in more detail in section 2.

On the other hand, \seb \hspace{0mm} extends \bpel \hspace{0mm} by featuring sessions as first class citizens.
A client wishing to interact with  a service begins by opening a session with this service. The set of possible interactions with the service forms the behaviour of the session.
In the present paper, we concentrate on the definition of untyped \seb \hspace{0mm}. In the future we plan to define \emph{session types} in order to allow us to verify interaction safety by means of type verification.

In order to define the \seb \hspace{0mm} language, we give a formal semantics that is a novel contribution in itself as it takes into account both the graph nature of the language and the static semantics that define how variables are declared and used, and this is applicable to other \bpel \hspace{0mm} like languages. Indeed, previous approaches either resort to process algebraic simplifications, thus neglecting control links which, in fact, are an essential part of \bpel; or are based on Petri nets and thus do not properly cover the static semantics that regulate the use of variables.

In our approach, the operational semantics is obtained in two steps. The first consists in the creation of what we have called a control graph. This graph takes into account the effect of the evaluation of the control flow and of join-conditions. Control graphs contain symbolic actions and no variables are evaluated in the translation into control graphs. The second step in the operational semantics describes the execution of services when they are part of an assembly made of a client and other Web services. Based on this semantics we formalize the concepts of interaction error and of interaction safety. These concepts will be further developed in future work on verification of session typing.

The rest of this paper is organized as follows. Section 2 provides an informal introduction to the \seb \hspace{0mm} language and contrasts its features with those of \bpel. Sections 3 and 4 give the syntax and semantics of untyped \seb. These sections are self-contained and do not require any previous knowledge of \bpel. Section 5 presents the semantics of networked service configurations described in \seb, and the concepts of interaction error and of interaction safety.
Relevant related work is surveyed in section 6 and the paper is concluded in section 7 along with a brief discussion on how session types might be used in \seb \hspace{0mm} in order to prove interaction safety.

\section{Informal introduction to \seb}
\label{informal}

%
{\bf \small Session initiation}. The main novelty in \seb, compared 
to \bpel, resides in the addition of the session initiation, 
a new kind of atomic activity, and in the way sessions impact the 
invoke and receive activities. The following is a typical
sequence of three atomic \seb \hspace{0mm} activities that can be performed by a client 
(we use a simplified syntax): $\ s@p; \ s!op_1(x); \ s?op_2(y).\ $
This sequence starts by a session initiation
activity, $s@p$, where $s$ is a session variable
and $p$ a service location variable (this corresponds to a \bpel \hspace{0mm} partnerlink). The execution of $s@p$
by the client and by the target service (the one whose
address is stored in $p$) has the following effects: 
(i) a fresh session id is stored in $s$, (ii) a new service
instance is created on the service side and is dedicated
to interact with the client, (iii) another fresh
session id is created on the service instance side and is bound 
to the one stored in $s$ on the client side.  
The second activity, $s!op_1(x)$, is the sending of
an invocation operation, $op_1$, with argument $x$.
The invocation is sent precisely to this 
newly created service instance.
The third activity of the sequence, $s?op_2(y)$,
is the reception of an invocation operation $op_2$
with argument $y$ that comes from this same service 
instance.
Note that invocation messages are all one
way and asynchronous: \seb \hspace{0mm} does not provide for synchronous
invocation.  
Furthermore  \seb \hspace{0mm} does not provide for
explicit correlation sets as does \bpel \hspace{0mm}.
But, on the other hand, sessions  are to be considered as implicit
correlation sets and, indeed, they can be emulated by them.
But, in this paper, we preferred to have sessions as 
an explicit primitive of the language so as to better discuss and illustrate their contribution. 
Hence, in \seb, session ids are the only means to identify 
source and target service instances. This is illustrated in 
the above example where the session variable 
$s$ is systematically indicated in the invoke 
and receive activities. 
Moreover, sessions involve two and only two partners 
and any message sent by one partner over a session 
is targeted at the other partner.
Biparty sessions are less expressive than correlation sets.
Nevertheless, at the end of the paper, we will give some ideas as to how this 
limitation can be lifted. \vspace{-3mm}\\

{\bf \small Structured activities}. \seb \hspace{0mm} has the 
principal structured activities of \bpel, i.e., flow, 
for running activities in parallel;
sequence, for sequential composition of activities, and pick,
which waits for multiple messages, the continuation behavior being dependent on the received
message. \seb \hspace{0mm} also inherits the possibility of 
having control links between different subactivities contained 
in a flow, as well as adding a join condition to any activity.
As in \bpel, a join condition requires that all its arguments
have a defined value (true or false) and must evaluate 
to true in order for the activity to be executable. 
\seb \hspace{0mm} also implements so-called dead path elimination
whereby all links outgoing from a canceled activity, 
or from its subactivities, have their values set 
to false.  \vspace{-3mm}\\

{\bf \small Imperative Computations}. Given that \seb \hspace{0mm} is a language
designed as a proof of concept, we wished to limit
its main features to interaction behavior. Hence,
imperative computation and branching are not part
of the language. Instead, they are assumed to be performed
by external services that can be called upon 
as part of the orchestration. This approach is similar 
to languages like Orc~\cite{orc} where the focus is on providing
the minimal constructs that allow one to perform service
orchestration functions and where imperative computation
and boolean tests are provided by external \emph{sites}. 
In particular, the original $do \ until$ iteration of \bpel \hspace{0mm} 
is replaced in \seb \hspace{0mm} by a structured activity called
``repeat'', given by the syntax: 
$(do  \ 
\text{{\sffamily pic}}_1
\ until \ 
\text{{\sffamily pic}}_2)$. The informal meaning of repeat is: 
perform $\text{{\sffamily pic}}_1$ repeatedly until the arrival of an 
invocation message awaited for in $\text{{\sffamily pic}}_2$. \vspace{-3mm}\\

\begin{figure}

\begin{center}
\includegraphics[scale=0.65]{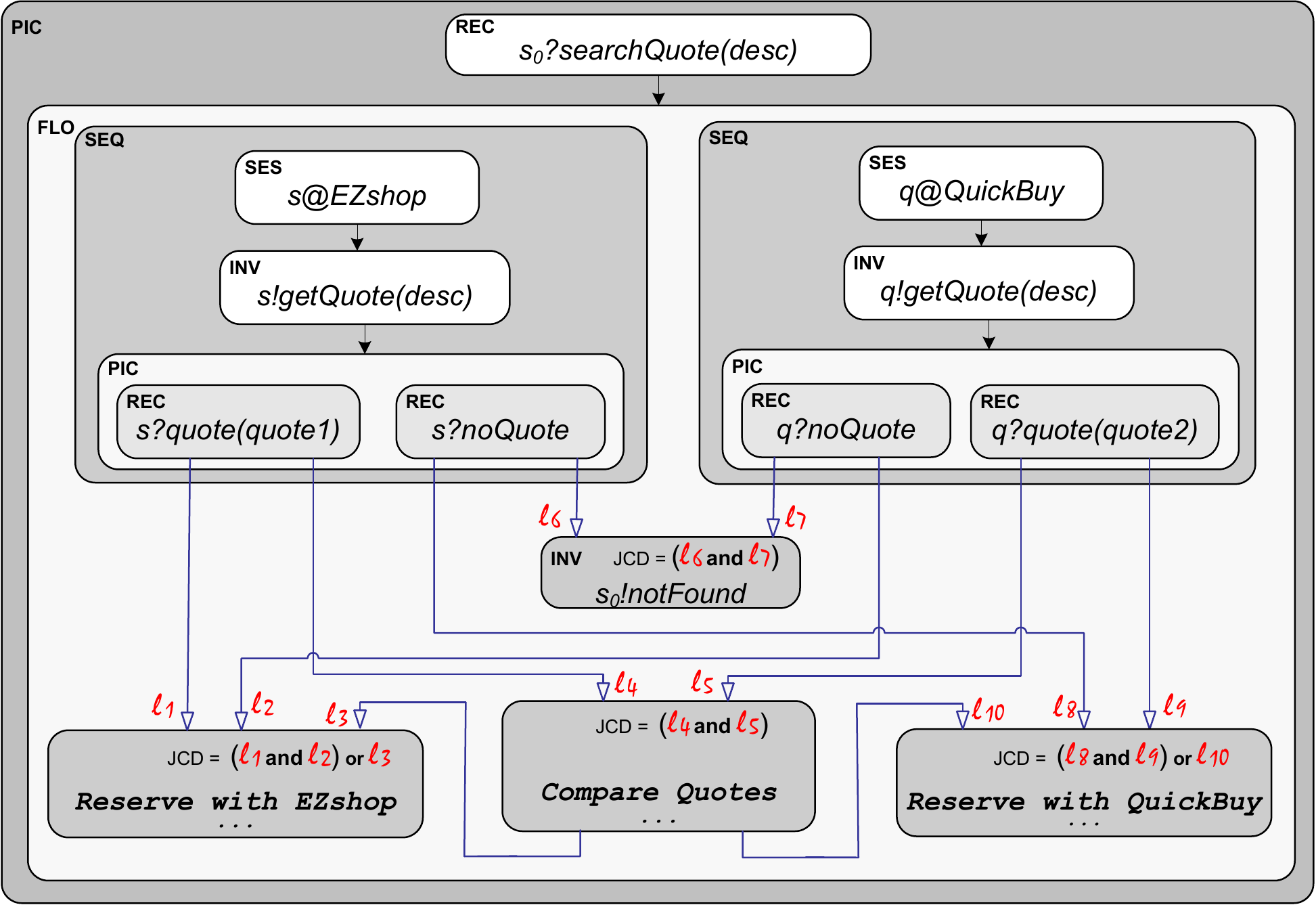}
\vspace{-3.5mm}
\caption{The QuoteComparer Service\label{quotecomparerdiag}}
\end{center}

\end{figure}

{\bf \small Example Service}. The QuoteComparer is an example of a service written in \seb \hspace{0mm} that will be used throughout the paper. Given an item description from a client, the purpose of the service is to offer a comparison of quotes from different providers for the item while at the same time reserving the item with the best offer. Figure \ref{quotecomparerdiag} contains a graphical representation of the QuoteComparer implementation. The representation is partial due to space constraints. The service waits for a client to invoke its $s_0?searchQuote$($desc$) operation with string parameter $desc$ containing an item description. Note the use of the special session variable $s_0$, called the root session variable. By accepting
the initial request from the client, the service implicitly begins an interaction with the client over session $s_0$.
The service then compares quotes for the item from two different providers ($EZshop$ and $QuickBuy$) by opening sessions with each of these providers.
Here, the sessions are explicitly opened: $s@EZshop$ is the opening of a session with the service having its
address stored in variable $EZshop$. The execution of $s@EZshop$ will result in a root session being initiated
at the $EZshop$ service and a fresh session id being stored in $s$. 

The service then behaves in the following way: depending on the returns made by the two providers, the QuoteComparer service either returns the best quote, the only available quote, or indicates to the client that no quote is available. The control links and join conditions illustrated in Figure \ref{quotecomparerdiag} implement this behaviour. For example, control link $l_1$ is set to true if $ EZshop$ returns message $quote(quote1)$, meaning $EZshop$ has an offer for the item. The value of this link and others will be later used to determine which provider's quote should be chosen. Indeed, the join condition (given by {\scshape jcd} = $(l_1 \; and \; l_2) \; or \; l_3$) of the bottom left box that deals with reserving an item with EZshop depends on the value of control link $l_1$. It also depends on control link $l_3$ which comes from the bottom central box that decides, in the case where both providers offer a quote, which quote is most advantageous. Link $l_2$ will be set to true only if provider $QuickBuy$ does not return a quote. Hence, the join condition of the bottom left box indicates that it should begin executing either if EZShop's quote is favourable
(the control link l$_3$ is set to true), or if EZShop returns a quote and QuickBuy returns $noQuote$ ($l_1$ and $l_2$ are set to true). The conditions for reserving with the $QuickBuy$ provider are symmetrical. 

If neither $QuickBuy$ nor $EZShop$ return a quote for the requested item, then control links $l_6$ and $l_7$ are set to true. This results in the execution of the central invocation operation $s_0!notFound$, i.e the client is informed that no quotes were found for the item description $desc$. \\

\vspace{-18mm}

\vspace{-2mm}

\section{Syntax of \seb~}

\vspace{-1.5mm}

\subsection{Basic Sets}

\seb \hspace{0mm} assumes three categories of basic sets: values, variables and others. They are introduced hereafter where, for each set,
a short description is provided as are the names of typical elements. All the sets are pairwise disjoint unless stated otherwise.

\begin{table}[!ht]
\centering
\renewcommand{\arraystretch}{1.05}
\begin{tabular}{| l | c | l | }
\hline
\multicolumn{1}{|c}{\textit{Set}} & 
\multicolumn{1}{|c}{\textit{Description}} & 
\multicolumn{1}{|c|}{\textit{Ranged over by}}  
\\
\hline\hline
$DatV\!al$ & Data Values & $ u, u', u_i, \ \cdots $  

 \\
\hline
$SrvV\!al $ & Service Locations & $\pi, \pi', \pi_i \ 
\cdots $ 
\\
\hline
$SesV\!al $ & Session Ids & $\alpha, \alpha', \alpha_i \ 
\cdots \ \beta \  \cdots $ 
\\
\hline
$ExV\!al$
& $\hspace{2mm}$ Exchangeable Values   $\hspace{2mm}$ & 
$w, w', w_i, \ \cdots$  
 \\
\hline
$LocV\!al$ & All Locations & 
$\delta, \delta', \delta_i, \ \cdots $ 
\\
\hline
$V\!al$ & All Values & 
$v, v', v_i, \ \cdots $ 
\\
\hline
\end{tabular}
 \vspace{-2mm} \caption{\label{values} \textbf{Values}}
\end{table}

Table \ref{values} presents the various sets of values used in \seb \hspace{0mm}. 
Data Values ($DatV\!al$), Service Locations 
($SrvV\!al $) and Sessions Ids ($SesV\!al $) are ground
sets. The set of Exchangeable Values $ExV\!al$ is given by 
$ExV\!al = DatV\!al \uplus SrvV\!al$, which means that 
both data values and service locations can be passed between services.
Hence, \seb \hspace{0mm} services may dynamically 
discover other services and may interact with them.  The set of all locations $LocV\!al$ is given by $LocV\!al= SrvV\!al \uplus SesV\!al$.
Hence, session ids are used to locate service instances.
The set of all values $V\!al$ is given by $V\!al= ExV\!al \uplus SesV\!al$.

\begin{table}[!ht]
\centering
\renewcommand{\arraystretch}{1.05}
\begin{tabular}{| l | c | l | }

\hline
\multicolumn{1}{|c}{\textit{Set}} & 
\multicolumn{1}{|c}{\textit{Description}} & 
\multicolumn{1}{|c|}{\textit{Ranged over by}}  
\\
\hline\hline
$DatV\!ar$ & Data Variables & $y$ \\
\hline
$SrvV\!ar$ & Service Location Variables & $p_0, p, p', p_i \ \cdots \ q \ \cdots$  \\
\hline
$ExV\!ar $ & \hspace{1mm} Variables of Exchangeable Values 
\hspace{1mm} & $x, x', x_i \ \cdots \ $ 
\\
\hline
$SesV\!ar $ & Session Variables & $s_0, s, s', s_i \ \cdots \ r \ \cdots $  
 \\
 \hline
$V\!ar $ & All Variables & $z, z', z_i \ \cdots \ $  
\\
\hline
\end{tabular}
 \vspace{-2mm} \caption{\label{variables} \textbf{Variables}}
\end{table}


Table \ref{variables} presents the various sets of variables. We can note 
two distinguished variables, $p_0$ and $s_0$:
$p_0$, is the service location variable that is
dedicated to holding a service's own location;  
$s_0$ is a session variable for accepting sessions
initiated by clients.
The use of $p_0$ and $s_0$  will be described in detail later on in the paper.
The set of variables of exchangeable values $ExV\!ar$  
and the set of all variables $V\!ar$ are defined by:
$ExV\!ar = DatV\!ar \uplus SrvV\!ar$ and
$V\!ar = ExV\!ar \uplus SesV\!ar$.

\begin{table}[!ht]
\centering
\renewcommand{\arraystretch}{1.05}
\begin{tabular}{| l | c | l | }
\hline
\multicolumn{1}{|c}{\textit{Set}} & 
\multicolumn{1}{|c}{\textit{Description}} & 
\multicolumn{1}{|c|}{\textit{Ranged over by}}  \\
\hline\hline

$O\;\!\!p$ & Operation Names & $op, op', op_i \ \cdots \ $  
\\
\hline
$Lnk$ & Control Links & $l, l', l_i \ \cdots \ $ 
\\
\hline
$2^{Lnk}$ & subsets of $Lnk$ & $L, L\! _{_S}, l_{_T} \ \cdots \ $ 
\\
\hline
$J\!cd$ & Join Conditions & $e, e', e_i \ \cdots \ f \ \cdots \ $ \\
\hline

\end{tabular}
 \vspace{-2mm} \caption{\label{other} \textbf{Miscellaneous Sets}} 
\end{table}

Table \ref{other} presents the other basic sets. $Lnk$ is the set of all control
links and $Jcd$ is the set of join conditions, i.e.,
boolean expressions over control links.
An example of a join condition: $e = (l_1 \ \text{and} \ l_2) \ \text{or} \ l_3$.

Reconsidering our previous example, $EZShop$ and $QuickBuy$ are service location variables
(elements of set $SrvV\!ar$, with values taken in the set of service locations $SrvV\!al$ ); 
$desc$ is a data variable (with values in the set $DatV\!al$); $s_0$, $s$ and $q$
are session variables (elements of $SesV\!ar$, taking their values 
in the set of session ids $SesV\!al$ ); 
$l_1$, $l_2$
and $l_3$ are control links; and finally, expression
{\emph {\small $\ (\l_3$  and $l_2$) or $\ l_3 \ $}}
is a join condition.

\vspace{-4mm}
\subsection{Syntax of \seb~ Activities}

\seb \hspace{0mm} being a dialect of \bpel, {\small XML} would have been the most appropriate 
metalanguage for encoding its syntax. However, for the purpose
of this paper, we have adopted a syntax based on records 
(\`a la Cardelli and Mitchell~\cite{records}) as it is better suited for 
discussing the formal semantics and properties of the language. 
By virtue of this syntax, all \seb \hspace{0mm} activities, except $\mynil$,
are records having the following predefined fields: 
$\text{{\footnotesize\scshape knd}}$ which identifies
the kind of the activity, $\text{{\footnotesize\scshape beh}}$,
which gives its behavior, $\text{{\footnotesize\scshape src}}$
(respectively $\text{{\footnotesize\scshape tgt}}$),
which contains a declaration of the set of control links for which the activity is the
source (respectively target), $\text{{\footnotesize\scshape jcd}}$
which contains the activity's join condition, i.e., a boolean expression
over control link names (those given in field 
$\text{{\footnotesize\scshape tgt}}$).
Moreover, the \emph{flow} activity has an extra
field, $\text{{\footnotesize\scshape lnk}}$, which contains
the set of links that can be used by the subactivities
contained in this activity. Field names are also used
to extract the content of a field from an activity, 
e.g., if {\sffamily act} is an activity, then 
$\text{{\sffamily act}}.\text{{\footnotesize\scshape beh}}$ yields
its behavior. For example: a \emph{flow} activity is given 
by the record  
$\langle \text{{\footnotesize\scshape knd}}=\flodec, 
\text{{\footnotesize\scshape beh}} = \textit{\small my\_behavior}, \  
\text{{\footnotesize\scshape tgt}}=L\hspace{-0.2mm}_{_{\text {\scshape t}}},
\text{{\footnotesize\scshape src}}= L\hspace{-0.2mm}_{_{\text {\scshape s}}},
\text{{\footnotesize\scshape jcd}}=e, 
\text{{\footnotesize\scshape lnk}} = L \rangle \ $
where $L\hspace{-0.2mm}_{_{\text {\scshape t}}}, \  L\hspace{-0.2mm}_{_{\text {\scshape s}}}$ and $L $ are sets of control link
names, and $e$ is a boolean expression over control link names.
Finally, for the sake of conciseness, we will often drop field 
names in records and instead we will associate a fixed position 
to each field. Hence, the
\emph{flow} activity given above becomes:
$\langle \flodec, 
\textit{\small my\_behavior}, \  
L\hspace{-0.2mm}_{_{\text {\scshape t}}},
L\hspace{-0.2mm}_{_{\text {\scshape s}}},
e, 
L \rangle$. \vspace{-3mm} \\


We let $  ACT $ be the set of all
activities and $\text{{\sffamily act}}$ a running element
of $ ACT $, the syntax of activities is 
given in the following table: \\
\vspace{-2mm}

\noindent
\begin{tabular}{l c l l l}

$\text{{\sffamily act}}$ & ::= & $\mynil$ & (* nil activity *)  \\
 & $|$ & $\text{{\sffamily ses}} \ \hspace{.5mm} 
   | \ \text{{\sffamily inv}} \ 
   | \ \text{{\sffamily rec}} $ & (* atomic activities *) \\
 & $|$ & $\text{{\sffamily seq}} \  
   | \ \text{{\sffamily flo}} \ \hspace{.7mm} 
   | \ \text{{\sffamily pic}} \ 
   | \ \text{{\sffamily rep}} \ \ $ & (* structured activities *) \\
\end{tabular}     
~ \vspace{3mm} \\
\noindent
\begin{tabular}{lllll}
$\text{{\sffamily ses}}$ & ::= & 
$\langle \sesdec, 
s@p, \  
L\hspace{-0.2mm}_{_{\text {\scshape t}}}, \
L\hspace{-0.2mm}_{_{\text {\scshape s}}}, \ 
e \rangle \ $ & & {\small (* session init *)} \\
$\text{{\sffamily inv}}$ & ::= & 
$\langle \invdec, 
s!op(x_1, \cdots, x_n), \  
L\hspace{-0.2mm}_{_{\text {\scshape t}}}, \
L\hspace{-0.2mm}_{_{\text {\scshape s}}}, \ 
e \rangle \ $ & & {\small (* invocation *) }    \\
$\text{{\sffamily rec}}$ & ::= & 
$\langle \recdec, 
s?op(x_1, \cdots, x_n), \  
L\hspace{-0.2mm}_{_{\text {\scshape t}}}, \
L\hspace{-0.2mm}_{_{\text {\scshape s}}}, \ 
e \rangle \ $ & & {\small (* reception *) } \\
$\text{{\sffamily seq}}$ & ::= & 
$\langle \seqdec, 
\text{{\sffamily act}}_1; \cdots; \text{{\sffamily act}}_n, \  
L\hspace{-0.2mm}_{_{\text {\scshape t}}}, \
L\hspace{-0.2mm}_{_{\text {\scshape s}}}, \ 
e \rangle \ $ &  &  {\small (* sequence *) } \\
$\text{{\sffamily flo}}$ & ::= & 
$\langle \flodec, 
\text{{\sffamily act}}_1 
| \cdots \ | 
\text{{\sffamily act}}_n, \  
L\hspace{-0.2mm}_{_{\text {\scshape t}}}, \
L\hspace{-0.2mm}_{_{\text {\scshape s}}}, \ 
e,
L
\rangle \ $ &  &  {\small (* flow *) } \\
$\text{{\sffamily pic}}$ & ::= & 
$\langle \picdec, 
\text{{\sffamily rec}}_1 ; \text{{\sffamily act}}_1
+ \cdots +
\text{{\sffamily rec}}_n ; \text{{\sffamily act}}_n, \  
L\hspace{-0.2mm}_{_{\text {\scshape t}}}, \
L\hspace{-0.2mm}_{_{\text {\scshape s}}}, \ 
e\rangle \ $ & &  {\small (* pick *) } \\
$\text{{\sffamily rep}}$ & ::= & 
$\langle \repdec, 
do  \ 
\text{{\sffamily pic}}_1
\ until \ 
\text{{\sffamily pic}}_2, \  
L\hspace{-0.2mm}_{_{\text {\scshape t}}}, \
L\hspace{-0.2mm}_{_{\text {\scshape s}}}, \ 
e\rangle \ $ & &  {\small (* repeat *) } \\
\end{tabular}
~ \vspace{1mm} \\

Note that in the production rule for 
$\text{{\sffamily flo}}$, 
``$|$'' is to be considered merely as a token separator. 
It is preferred over comma because it is more 
visual and better conveys the intended intuitive meaning 
of the $\text{{\sffamily flo}}$ activity, which is to be the
container of a set of sub activities that
run in parallel.
The same remark applies to symbols ``;'', ``$+$'',
``$do$'' and ``$until$'' which are used as token separators  
in the production rules for $\text{{\sffamily seq, pic}} $
and $\text{{\sffamily rep}}$ to convey their appropriate 
intuitive meanings. Note that ``$|$'' and ``$+$'' are commutative.
As a final note, the number $n$ appearing in the rules for $\text{{\sffamily seq}}$, 
$\text{{\sffamily flo}}$ and $\text{{\sffamily pic}}$
is such that $ n \geq 1 $.  \vspace{-3mm} \\

\noindent
Returning to Figure \ref{quotecomparerdiag}, the syntax of the central $\invdec$ activity is given by the following expression: 
$\ \ \ \ \langle \invdec, \ s_0!notFound , \ \{l_6,l_8\}, \ \emptyset, \ l_6 \ and \ l_8 \rangle, \ \; $
and the syntax of the $\text{{\sffamily seq}}$ activity at the top left of the example is given by the following 
expression: \\

\vspace{-2mm}
\noindent
\begin{tabular}{ll}
$\ \ \ $  $\langle \seqdec, \ $ & \vspace{-2mm} \\ 
$\ \ \ $  & $\langle \sesdec, s@EZshop, \ \emptyset, \ \emptyset, \ true \rangle \ ;$ \\
$\ \ \ $  & $\langle \invdec \!, s!getQuote(desc), \ \emptyset, \ \emptyset, \ true \rangle \ ;$ \\
$\ \ \ $  & $\langle \picdec, $ \\
$\ \ \ $  & $\ \ \ \ \ \ \ \ \langle \recdec \!, s?quote(quote1), \ \emptyset, \ \{l_1,l_4\}, \ true \rangle \ + \ \langle \recdec \!, s?noQuote, \ \emptyset, \ \{l_6,l_8\}, \ true \rangle , $ \\
$\ \ \ $  & $ \ \emptyset , \emptyset, \ true \rangle,$ \vspace{-2mm} \\  
$\ \ \ $  $\ \ \emptyset, \emptyset, \ true \rangle $ & \vspace{2.5mm} \\
\end{tabular}

\noindent
\textbf{Note: Syntax simplification} - when  an activity 
has no incoming or outgoing control links, we may omit 
its encapsulating record and represent it just by
its behavior. Hence, for example, we may write $\text{\sffamily act};\text{\sffamily act'}$
instead of 
$\langle \seqdec, 
\text{{\sffamily act}}; \text{{\sffamily act'}}, \  
\emptyset, \
\emptyset, \ 
true \rangle$, and we will write $s!op(x)$ instead of
$\langle \invdec, 
s!op(x), \  
\emptyset, \
\emptyset, \ 
true \rangle$. 

\begin{mydef} \emph{Subactivities} - For an activity $\text{\sffamily{act}}$, 
we define the sets of activities, $\ \widehat{\text{\sffamily{act}}}  \ $ 
and $\ \widehat{\widehat{\text{\sffamily{act}}}}$:
\vspace{-1mm}
\begin{itemize}
\item
$\widehat{\text{\sffamily{act}}} \ \stackrel{\Delta}{=} \ 
\{ \text{\sffamily{act}} \}$ if $\ \text{\sffamily{act}} \ $ is an atomic activity, else
$\widehat{\text{\sffamily{act}}} \ \stackrel{\Delta}{=} \ 
\{ \text{\sffamily{act}} \} \ \cup \ \widehat{\text{\sffamily{act}}.\text{{\footnotesize\scshape beh}}}$
\vspace{-1mm}
\item
$\widehat{\widehat{\text{\sffamily{act}}}} \ \stackrel{\Delta}{=} \ 
\emptyset \ $
if $\ \text{\sffamily{act}} \ $ is an atomic activity, else
$\widehat{\widehat{\text{\sffamily{act}}}} \ \stackrel{\Delta}{=} \ 
\widehat{\text{\sffamily{act}}.\text{{\footnotesize\scshape beh}}}$

\end{itemize}
\end{mydef}

\vspace{-1mm}

\begin{quote}
\textit{Informally, $\ \widehat{\text{\sffamily{act}}}  \ $  is the set of activities
transitively contained in $\ \text{\sffamily{act}} \ $ 
and $\ \widehat{\widehat{\text{\sffamily{act}}}} \ $ is the set of activities
strictly and transitively contained in $\ \text{\sffamily{act}}$}.
\end{quote}

\begin{mydef} \emph{Precedence relation between activities} - For an activity $\text{\sffamily{act}}$, 
we define relation $pred$ on the set $\widehat{\text{\sffamily{act}}}$ as follows: \\
$\text{{\sffamily act}}_1 
\ pred \ 
\text{{\sffamily act}}_2 \ \ \ $ iff 
$\ \ (\ \text{{\sffamily act}}_1.{\text{\footnotesize\scshape src}} \cap
\text{{\sffamily act}}_2.{\text{\footnotesize\scshape tgt}} \neq \emptyset \ ) \ $ or 
$\ (\ \exists
\text{{\sffamily seq}} \in \widehat{\text{{\sffamily act}}} 
\ \text{with} \ 
\text{{\sffamily seq}}.{\text{\footnotesize\scshape beh}}=
 \cdots
\text{{\sffamily act}}_1;
\text{{\sffamily act}}_2
\cdots \ )  
$ \ \ \ \ \ \
\end{mydef}

\vspace{-1mm}

\begin{quote}
\textit{Informally, relation $pred$ implies that  $\ {\text{\sffamily{act}}}_1  \ $  
preceeds $\ {\text{\sffamily{act}}}_2  \ $ either in some
\text{{\sffamily seq}} activity or because $\ {\text{\sffamily{act}}}_1  \ $
has at least one outgoing control link targeting $\ {\text{\sffamily{act}}}_2$}.
\end{quote}

Activities need to comply to certain constraints concerning the declaration
and usage of their control links. Firstly, 
control links should be well scoped, unique and should form
no cycles. 
Secondly,  
all $\text{\sffamily{rep}}$ subactivities
should be well-formed in terms of incoming and outgoing control links.
An activity that complies with these constraints is said to be well-formed. \vspace{1mm}

\begin{mydef}\emph{$\text{\sffamily{rep}}$-well-formed activity} -
A \seb \hspace{0mm} activity $\text{\sffamily{act}}_0$ is $\text{\sffamily{rep}}$-well-formed $i\!f\!\!f$ 
any activity  \\ $\text{\sffamily{rep}}=
\langle \repdec, 
do  \ 
\text{{\sffamily pic}}_1
\ until \ 
\text{{\sffamily pic}}_2, \  
L\hspace{-0.2mm}_{_{\text {\scshape t}}}, \
L\hspace{-0.2mm}_{_{\text {\scshape s}}}, \ 
e\rangle$ of $ \ \widehat{\text{\sffamily{act}}_0} \ $  
satisfies the following 3 conditions:

\noindent
(1)  $\text{{\sffamily pic}}_1.\text{{\scshape \footnotesize tgt}}= 
\text{{\sffamily pic}}_2.\text{{\scshape \footnotesize tgt}} = \emptyset $, \ \ 
(2) $\ \text{{\sffamily pic}}_1.\text{{\scshape \footnotesize src}} = \emptyset$,
 \ \ and \ (3) \ $\widehat{\widehat{\text{{\sffamily pic}}}}_1.\text{{\scshape \footnotesize src}}=
\widehat{\widehat{\text{{\sffamily pic}}}}_1.\text{{\scshape \footnotesize tgt}}$.

\end{mydef}
\begin{quote}
\textit{Informally, (1)  implies that $\text{{\sffamily pic}}_1$ and $\text{{\sffamily pic}}_2$ 
have no incoming links, (2) states that $\text{{\sffamily pic}}_1$ has no outgoing links and (3) states that
all control links of activities (strictly) contained in $\text{{\sffamily pic}}_1$
are (strictly) internal to $\text{{\sffamily pic}}_1$.}
\end{quote}
\begin{mydef} \emph{Well-structured activity} - 
A \seb \hspace{0mm} activity $\text{\sffamily{act}}_0$ is well structured $if\!f$ the control links
occurring in any activity of $\widehat{\text{\sffamily{act}}_0}$ satisfy the unicity, 
scoping and non cyclicity
conditions given below, where $\text{\sffamily{act}}$, $\text{\sffamily{act}}'$, $\text{\sffamily{act}}''$ 
and $\text{\sffamily{seq}}$ are subactivities of $\text{\sffamily{act}}_0$.

\vspace{-1mm}

\begin{enumerate}

\item \emph{Control links unicity} -
\noindent Given any control link $l$, and any pair of activities {\sffamily act} and {\sffamily act'}: \\
\noindent
if 
($l  \in$ {\sffamily act}.{\footnotesize\scshape lnk} 
$\cap$ 
{\sffamily act'}.{\footnotesize\scshape lnk})
or 
($l  \in$ {\sffamily act}.{\footnotesize\scshape src} 
$\cap$ 
{\sffamily act'}.{\footnotesize\scshape src})
or
($l  \in$ {\sffamily act}.{\footnotesize\scshape tgt} 
$\cap$ 
{\sffamily act'}.{\footnotesize\scshape tgt}) then {\sffamily act} = {\sffamily act'}.

\vspace{-1mm} 
 
\item \emph{Control links scoping} -
\noindent If $l \in \text{{\sffamily act}}.{\text{\footnotesize\scshape src}}$
(respectively if
$l \in \text{{\sffamily act}}.{\text{\footnotesize\scshape tgt}} \ $)
then
$\ \exists \ \text{{\sffamily act'}}, \text{{\sffamily act''}}$ 
with {\sffamily act} $\in$ $\widehat{\text{\sffamily act''}}$ 
and 
$\text{\sffamily act'} \in$ $\widehat{\text{\sffamily act''}}$ and 
with $l \ \in \text{\sffamily act''}.\text{\footnotesize\scshape lnk} \ $ 
and $l \ \in \text{\sffamily act'}.\text{\footnotesize\scshape tgt}$. 
(respectively $l \ \in \text{\sffamily act'}.\text{\footnotesize\scshape src}$).

\vspace{-1mm}

\item \emph{Control links non cyclicity}:\\
\noindent 
(i) Relation $pred$ is acyclic, and \\
(ii) $\forall \text{{\sffamily act}}, \text{{\sffamily act'}} \in \widehat{\text{\sffamily{act}}_0} 
\ \ \ \  
\text{{\sffamily act'}} \in  \widehat{\text{\sffamily{act}}} 
\ \ \ \Rightarrow  
\ \ (\ \text{{\sffamily act}}.{\text{\footnotesize\scshape src}} \cap
\text{{\sffamily act}}'.{\text{\footnotesize\scshape tgt}} = \emptyset \ ) \  \text{and} \ \ 
(\ \text{{\sffamily act}}'.{\text{\footnotesize\scshape src}} \cap
\text{{\sffamily act}}.{\text{\footnotesize\scshape tgt}} = \emptyset \ )$ 
\end{enumerate}
\end{mydef}

\vspace{-1mm}

\begin{quote}
\textit{Informally, a well-structured activity is such that
all of its control links (including those in subactivities) are well scoped (i.e., are within the scope of one flow subactivity),
unique (each target declaration corresponds to one and only 
one source declaration and vice versa), and do not form 
any causality cycles, either directly (from source to target of activities) 
or through activities that are chained within some sequence activity 
or through the containment relation between activities.}
\end{quote}

\begin{mydef} \emph{Well-formed activity} -
an activity is well formed $i\!f\!\!f$  
it is both well structured and $\text{\sffamily{rep}}$-well-formed.
\end{mydef}

\vspace{-5mm}

\section{Semantics of \seb~}

\vspace{-1mm}

Here we define the notion of control graphs and then provide the \sos \hspace{0mm} rules that define a translation from a well-formed activity
into a control graph. We then study the properties of control graphs. The semantics of \seb\hspace{0mm} activities is obtained by applying a series of transformations to this first control graph that lead to a final control graph,
noted $\textbf{cg}(\text{{\sffamily act}})$.


\vspace{-2mm}

\subsection{Definitions}

\vspace{-1mm}

\subsubsection{Control Graphs}

\vspace{-1mm}

\begin{mydef} \emph{Observable Actions} - The set, \emph{$\text{\footnotesize \sffamily{ACTIONS}}$}, 
of  \textit{observable} actions is defined by: \\
$\ \ \emph{\text{\footnotesize \sffamily{ACTIONS}}} =_{\text{\tiny def}} \{ \ \ a \ \ | \ \
a \  \text{is any action of the form:} \ \ \ 
s@p, \ \ s!op(x_1, \cdots, x_n)  \ \ \ \text{or} \ \ \ s?op(x_1, \cdots, x_n) \ \ \}$ 
\end{mydef}

\begin{mydef} \emph{All actions} - We define the set $\emph{\text{\footnotesize \sffamily{ACTIONS}}}_{\tau}$
of all actions (ranged over by $\sigma$):
$\emph{\text{\footnotesize \sffamily{ACTIONS}}}_{\tau} \! =_{\text{\tiny def}}
\emph{\text{\footnotesize \sffamily{ACTIONS}}} \cup \{\tau\} $ where  
$\tau$ denotes the unobservable (or silent) action. 
\end{mydef}

\begin{mydef} \emph{Control Graphs} - A control graph, $\Gamma = <G,g\!\!\;_{_{\scriptscriptstyle{0}}},{\cal A} ,\rightarrow > \ $ , is a labelled transition system where: \ \
\begin{tabular}{lll}
$\ \ \ \ \ $ & $-\ G $ is a set of states, called control states
&
$\ \ \ - \ g\!\!\;_{_{\scriptscriptstyle{0}}} $ is the initial control state 
\\
$\ \ \ \ \ $ & $- \ {\cal A} \ $ is a set of actions $ ({\cal A} \subset 
\emph{\text{\footnotesize \sffamily{ACTIONS}}}_{\tau})$ 
&
$\ \ \ - \ \rightarrow \ \subset \ G \ \times \ {\cal A} \ \times \ G $ 
\end{tabular}
\end{mydef}

\subsubsection{Control Link Maps}


We now define the control part of activities where we consider only the
values of control links. Activities are given a map that stores the running values
of control links, which are initially undefined values.

\begin{mydef} - \emph{Control Link Maps} - A Control link map $c$ is a partial function from the set of control links, $Lnk$, to 
the set of boolean values extended with the undefined value.
$\ \ \ c: Lnk \rightarrow \{ 
\text{{\footnotesize{\bf true}}},
\text{{\footnotesize{\bf false}}},
\bot \}$
\end{mydef}

\begin{mydef} \emph{Initial Control Links Map} - For an activity $\text{\sffamily{act}}$ we define
$c_\text{\sffamily{act}}$, the initial
control links map: $ \ \ dom(c_\text{\sffamily{act}})= \{ \ l \in  Lnk \ | \ l \ \text{occurs in}   
\ \widehat{\text{\sffamily{act}}} \} $  
\; and \; $\forall \ \! l \in dom(c_\text{\sffamily{act}}): c_\text{\sffamily{act}}(l)= \perp$ 
\end{mydef}

\begin{mydef} \emph{Evaluation of a Join Condition} - If $L$ is a set of control links, $e$ a boolean expression over $L$ and $c$ a control links map,
then the evaluation of $e$ in the context of $c$ is written: $c \triangleright e(L)$. 
Furthermore, we consider that this evaluation is defined only when 
$\forall l \in L, c(l) \neq \bot$.
\end{mydef}

\subsection{From Activities to Control Graphs: Structured Operational Semantics Rules}

\label{semantics}




\begin{table}[!ht]
\noindent
\begin{minipage}{16.2cm}
\fontsize{9.5pt}{4pt}
\selectfont
\hspace{-2mm}
\begin{minipage}{3.5cm}
\regle{
$\fbox{\textcolor{black}{{\tiny SES}}} \ \ \ \ $
\hspace{-2mm}
 c \triangleright e(L\hspace{-0.1mm}_{_{\text {\scshape t}}}) =
\text{{\footnotesize{\bf true}}} \hspace{-2mm}
\vspace{1mm}

}
{ \vspace{-1mm} \\
\hspace{-2mm}
(c, \langle \!\; \sesdec , s@p \;\! ,
L\hspace{-0.1mm}_{_{\text {\scshape t}}} \hspace{.2mm} ,
L\hspace{-0.2mm}_{_{\text {\scshape s}}} \hspace{.2mm} ,
e \hspace{.2mm}\rangle ) \vspace{1mm} \\
\ \ \ \ \ \ \ \ \ \ \ \ \ \ \ \ \downarrow
{\scriptstyle s@p} \vspace{1mm} \\
\ \ \ \ \ \ \ \ (c
[
\hspace{0.4mm}
^{\text{{\footnotesize{\bf true}}}}
\hspace{-0.7mm}
/
\hspace{-0.7mm}
_{L\hspace{-0.2mm}_{_{\text {\scshape s}}}}
]
\; ,
 \mynil )
}
\end{minipage}
%
\hspace{1mm}
%
\begin{minipage}{3.9cm}
\regle{
$\fbox{\textcolor{black}{{\tiny INV}}} \ \ \ \ $
c \triangleright e(L\hspace{-0.1mm}_{_{\text {\scshape t}}}) =
\text{{\footnotesize{\bf true}}}
\vspace{1mm}

}
{ \vspace{-1mm} \\
\hspace{-2mm}
(c, \langle \!\; \invdec , s!op(\tilde{x}) \;\! ,
L\hspace{-0.1mm}_{_{\text {\scshape t}}} \hspace{.2mm} ,
L\hspace{-0.2mm}_{_{\text {\scshape s}}} \hspace{.2mm} ,
e \hspace{.2mm}\rangle ) \vspace{1mm} \\
\ \ \ \ \ \ \ \ \ \ \ \ \ \ \ \ \ \ \downarrow
{\scriptstyle s!op(\tilde{x})} \vspace{1mm} \\
\ \ \ \ \ \ \ \ (c
[
\hspace{0.4mm}
^{\text{{\footnotesize{\bf true}}}}
\hspace{-0.7mm}
/
\hspace{-0.7mm}
_{L\hspace{-0.2mm}_{_{\text {\scshape s}}}}
]
\; ,
 \mynil )

}
\end{minipage}
%
%
\hspace{1mm}
%
%
\begin{minipage}{3.9cm}
\regle{
$\fbox{\textcolor{black}{{\tiny REC}}}$
\hspace{5mm}
c \triangleright e(L\hspace{-0.1mm}_{_{\text {\scshape t}}}) =
\text{{\footnotesize{\bf true}}} \

\vspace{1mm}

}
{ \vspace{-1mm} \\
\hspace{-1mm}
(c, \langle \!\; \recdec , s?op(\tilde{x}) \;\! ,
L\hspace{-0.1mm}_{_{\text {\scshape t}}} \hspace{.2mm} ,
L\hspace{-0.2mm}_{_{\text {\scshape s}}} \hspace{.2mm} ,
e \hspace{.2mm}\rangle ) \vspace{1mm} \\
\ \ \ \ \ \ \ \ \ \ \ \ \ \ \ \ \ \ \downarrow
{\scriptstyle s?op(\tilde{x})} \vspace{1mm} \\
\ \ \ \ \ \ \ \ (c
[
\hspace{0.4mm}
^{\text{{\footnotesize{\bf true}}}}
\hspace{-0.7mm}
/
\hspace{-0.7mm}
_{L\hspace{-0.2mm}_{_{\text {\scshape s}}}}
]
\; ,
 \mynil )

}
\end{minipage}
%
%
%
\hspace{3mm}
%
%
\begin{minipage}{3.7cm}
\fontsize{9pt}{10pt}
\selectfont
%
\
%
\vspace{-2mm}
\ \ \ \ \ \ \ \ \ \ \ \ \ \ \ \ \ \ \ \ \ \ \ \ \ \ \ \regle{
$ \fbox{{\tiny DPE}} \ \ \ \ $
c \triangleright e(L\hspace{-0.1mm}_{_{\text {\scshape t}}}) =
\text{{\footnotesize{\bf false}}}
\vspace{1mm}
}
{
(c, \langle \anydec, \; \text{{\sffamily act}}, \;
L\hspace{-0.1mm}_{_{\text {\scshape t}}}, \;
L\hspace{-0.2mm}_{_{\text {\scshape s}}}, \; e \rangle)
 \\
\ \ \ \ \ \ \ \ \
\ \ \ \ \ \ \ \
 \downarrow{ \! \tau} \\
(c
[ \hspace{.3mm}
^{\text{{\footnotesize{\bf false}}}}
\hspace{-0.7mm} / \hspace{-0.4mm}
_{\widehat{\text{\footnotesize \sffamily{act}}}.\text{{\scshape \footnotesize src}}}
\hspace{.3mm} ]
, \;
  \mynil \ )
}

\end{minipage}
\end{minipage}

\vspace{0.4cm}

\begin{minipage}{16.2cm}
\vspace{-2.5mm}

\fontsize{9pt}{10pt}
\selectfont
\begin{minipage}{7.5cm}
\begin{center}
\vspace{4mm}
\regle{ 
\hspace{-0.5cm} $ \fbox{{\tiny PIC}} $ 
\ \ \ \ \ \ \ \ \ c \triangleright e(L\hspace{-0.1mm}_{_{\text {\scshape t}}})\!=\!
\text{{\footnotesize{\bf true}}}
\ \ \
(c,
 \text{\sffamily{rec}})
\!\!\!\!\! \Fait{\sigma} \!\!\!\!\!
(c', \mynil )
\vspace{1mm}
}
{
\vspace{-3mm} \\
\ \ \ \ \ \ \ \ \ \ \ (\ c, \
\langle \picdec ,
\text{\sffamily{rec}};\text{\sffamily{act}}  
+
\sum \text{\sffamily{rec}}_i ; \text{\sffamily{act}}_i
\; , \; L\hspace{-0.1mm}_{_{\text {\scshape t}}}, \;
L\hspace{-0.2mm}_{_{\text {\scshape s}}}, \; e \ \rangle \
 ) \\
\ \ \ \ \ \ \ \ \ \ \ \ \ \ \ \ \ \ \ \ \ \ \ \ \ \ \ \ \ \ \ \ \ \ \ \downarrow \! \sigma \\
(\ c'[ \hspace{.3mm}
^{\text{{\footnotesize{\bf false}}}}
\hspace{-0.7mm} / \hspace{-0.2mm}
\widehat{
(\sum \text{\sffamily{rec}}_i ; \text{\sffamily{act}}_i ).\text{{\scshape \footnotesize src}} }
\hspace{.3mm} ], \ \langle
\flodec ,   \
\text{\sffamily{act}} \;
, \; L\hspace{-0.1mm}_{_{\text {\scshape t}}}, \;
L\hspace{-0.2mm}_{_{\text {\scshape s}}}, \; e \;
\rangle
\; ) \vspace{3mm} \\
}
\end{center}
\end{minipage}
%
\begin{minipage}{8.2cm}
\vspace{-0.5mm}
\begin{center}
\regle{ \hspace{37mm} 
\vspace{-3mm}\\
$\fbox{{\tiny REP1}}  $
\; c \triangleright e(L\hspace{-0.1mm}_{_{\text {\scshape t}}}) = 
\text{{\footnotesize{\bf true}}}
\
(c, \text{\sffamily{pic}}_1)
\fait{\sigma}
(c', \text{\sffamily{act}})
\vspace{1mm}
}
{
\ \ \ \ \ (c, \; \langle \repdec, \;
\! \textit{do} \;  \text{\sffamily{pic$_1$}} \;\! \textit{until} \; \text{\sffamily{pic$_2$}} , \;
L\hspace{-0.1mm}_{_{\text {\scshape t}}}, \;
L\hspace{-0.2mm}_{_{\text {\scshape s}}},
\; e \rangle \; ) \\
\ \ \ \ \
\ \ \ \ \ \ \ \ \ \ \ \ \ \ \ \ \ \ \ \ \ \ \  \downarrow{ \! \sigma} \\
(c' \! , \! \langle \unfdec, \textit{do} \; \text{\sffamily{act}} \; \textit{then} \; \text{\sffamily{pic$_1$}} \;\! \textit{until} \; \text{\sffamily{pic$_2$}}, \;\! 
L\hspace{-0.1mm}_{_{\text {\scshape t}}}, \;\!
L\hspace{-0.2mm}_{_{\text {\scshape s}}}, \;\! e \rangle ) \\

}

\end{center}
\end{minipage}
\end{minipage}

\vspace{0.4cm}

\begin{minipage}{16.2cm}

\fontsize{9pt}{10pt}
\selectfont
\vspace{-2mm}
\begin{minipage}{7.5cm}
\begin{center}

\begin{center}
\regle{
\vspace{-3mm}\\
$\fbox{{\tiny FLO1}} $  \hspace{0.5cm}
c \triangleright e(L\hspace{-0.1mm}_{_{\text {\scshape t}}}) \!
= \!  \text{{\footnotesize{\bf true}}}
\ \ \ \
(c,
 \text{\sffamily{act}}_i)
\fait{\sigma}
(c', \text{\sffamily{act}}')
\vspace{1mm}
}
{
\vspace{-3mm} \\
(c,
\langle \flodec \!,  \
\text{\sffamily{act}}_1
| {\scriptscriptstyle \cdots } | \text{\sffamily{act}}_i |
{\scriptscriptstyle \cdots } |
\text{\sffamily{act}}_n \;
, \;
L\hspace{-0.1mm}_{_{\text {\scshape t}}}, \;
L\hspace{-0.2mm}_{_{\text {\scshape s}}}, \; e \rangle
 ) \\
\ \ \ \ \ \ \ \ \ \ \ \ \ \ \ \ \ \ \ \
\downarrow \! \sigma \ \ \ \ \ \ \ \ \\
(c',  \langle
\flodec \!,   \
\text{\sffamily{act}}_1
| {\scriptscriptstyle \cdots } | \text{\sffamily{act}}' |
{\scriptscriptstyle \cdots } |
\text{\sffamily{act}}_n \;
, \;
L\hspace{-0.1mm}_{_{\text {\scshape t}}}, \;
L\hspace{-0.2mm}_{_{\text {\scshape s}}}, \; e
\rangle
)
}
\end{center}

\end{center}
\end{minipage}
%
\begin{minipage}{8.2cm}
\begin{center}

\begin{center}
\regle{
\hspace{-1.5mm}$\fbox{{\tiny REP2}} $ \hspace{0.1cm}
 c \triangleright e(L\hspace{-0.1mm}_{_{\text {\scshape t}}}) \!=\!
\text{{\footnotesize{\bf true}}}
\
(c, \text{\sffamily{pic}}_2)
\fait{\sigma}
(c', \text{\sffamily{act}})
\vspace{1mm}
}
{
(c, \; \langle \repdec,
\; \textit{do} \;  \text{\sffamily{pic$_1$}} \;\! \textit{until} \; \text{\sffamily{pic$_2$}} , \;
L\hspace{-0.1mm}_{_{\text {\scshape t}}}, \;
L\hspace{-0.2mm}_{_{\text {\scshape s}}},
\; e \rangle \; ) \\
\ \ \ \ \ \ \ \ \ \ \ \ \
\ \ \ \ \ \ \ \ \ \ \ \ \downarrow{ \! \sigma} \\
\ \ \ \ \ \ \ \ (c', \; \langle \flodec, \; \text{\sffamily{
 act }}, \;
L\hspace{-0.1mm}_{_{\text {\scshape t}}}, \;
L\hspace{-0.2mm}_{_{\text {\scshape s}}},
\; e \rangle \; ) \\
}
\end{center}

\end{center}
\end{minipage}
\end{minipage}

\vspace{0.4cm}

\begin{minipage}{16.2cm}

\fontsize{9pt}{10pt}
\selectfont
\begin{minipage}{7cm}
\begin{center}
\vspace{3mm}

\regle{
\hspace{-.7cm} \fbox{{\tiny FLO2}} 
\ \ \ \ \ \ \ \ \ \ c \triangleright e(L\hspace{-0.1mm}_{_{\text {\scshape t}}})
 = \text{{\footnotesize{\bf true}}}
\ \ \ \ \ \ \ \ \
\vspace{1mm}
}
{
\vspace{-3mm} \\
(\ c, \
\langle \flodec \!,
\text{\sffamily{act}}_1
| {\scriptscriptstyle \cdots }
| \mynil \hspace{0.1mm}
| {\scriptscriptstyle \cdots }
| \text{\sffamily{act}}_n \;
, \;
L\hspace{-0.1mm}_{_{\text {\scshape t}}}, \;
L\hspace{-0.2mm}_{_{\text {\scshape s}}}, \; e \rangle \
 ) \\
\ \ \ \ \ \ \ \ \ \: \ \ \ \ \ \ \ \ \ \ \ \ \ \ \downarrow \! \tau \\
(\ c, \ \langle
\flodec \!,   \
\text{\sffamily{act}}_1
| {\scriptscriptstyle \cdots }
| {\scriptscriptstyle \cdots }
|
\text{\sffamily{act}}_n \;
, \;
L\hspace{-0.1mm}_{_{\text {\scshape t}}}, \;
L\hspace{-0.2mm}_{_{\text {\scshape s}}}, \; e
\rangle
\; )
}

\begin{center}

\end{center}

\end{center}
\end{minipage}
%
\begin{minipage}{8.7cm}
\begin{center}

\vspace{-2.5mm}
\begin{center}

\regle{
\vspace{-4mm}\\
$\fbox{{\tiny UNF1}} \hspace{0.1cm}$
c \triangleright e(L\hspace{-0.1mm}_{_{\text {\scshape t}}}) =
\text{{\footnotesize{\bf true}}}
\ \
(c, \text{\sffamily{act}})
\fait{\sigma}
(c', \text{\sffamily{act'}}) 
\vspace{1mm}
}
{
\vspace{-3mm} \\
(c, \langle \unfdec, \textit{do} \; \text{\sffamily{act}} \; \textit{then} \; \text{\sffamily{pic$_1$}} \;\! \textit{until} \; \text{\sffamily{pic$_2$}} , \;
L\hspace{-0.1mm}_{_{\text {\scshape t}}}, \;
L\hspace{-0.2mm}_{_{\text {\scshape s}}},
\; e \rangle) \\
\ \ \ \ \ 
\ \ \ \ \ \ \ \ \ \ \ \ \ \ \ \ \ \ \ \ \ \ \  
\downarrow{ \! \sigma} \\
(c', \langle \unfdec, \textit{do} \; \text{\sffamily{act'}} \; \textit{then} \; \text{\sffamily{pic$_1$}} \;\! \textit{until} \; \text{\sffamily{pic$_2$}}, \; 
L\hspace{-0.1mm}_{_{\text {\scshape t}}}, \;
L\hspace{-0.2mm}_{_{\text {\scshape s}}},
\; e \rangle) \\

}

\end{center}

\end{center}
\end{minipage}
\end{minipage}

\vspace{-0.34cm}

\begin{minipage}{16.2cm}

\fontsize{9pt}{10pt}
\selectfont
\begin{minipage}{7.5cm}
\begin{center}

\vspace{-3mm}
\begin{center}
\vspace{-1mm}
\regle{
\hspace{-12mm} \fbox{{\tiny FLO3}}
\ \ \ \ \ \ \ \ \ c \triangleright
e(L\hspace{-0.2mm}_{_{\text {\scshape t}}})
 = \text{{\footnotesize{\bf true}}}
 \ \ \ \
\vspace{1mm}
}
{
(c, \langle \flodec \!, \; \mynil, \;\!
L\hspace{-0.1mm}_{_{\text {\scshape t}}}, \;\!
L\hspace{-0.2mm}_{_{\text {\scshape s}}}, \; e \rangle)          
           \fait{\tau}
(c
[
\hspace{0.4mm}
^{\text{{\footnotesize{\bf true}}}}
\hspace{-0.7mm}
/
\hspace{-0.7mm}
_{L\hspace{-0.2mm}_{_{\text {\scshape s}}}}
]
 , \mynil ) \hspace{-2mm}

}
\end{center}

\end{center}
\end{minipage}
%
\begin{minipage}{8.2cm}
\begin{center}

\begin{center}
\regle{
\vspace{0mm}\\
$\fbox{{\tiny UNF2}} \hspace{5cm} \vspace{-4mm}$\\ 
c \triangleright e(L\hspace{-0.1mm}_{_{\text {\scshape t}}}) =
\text{{\footnotesize{\bf true}}}
\vspace{1mm}
}
{
\vspace{-3mm} \\
(c, \;\! \langle \unfdec, \textit{do} \; \mynil \; \textit{then} \; \text{\sffamily{pic$_1$}} \;\! \textit{until} \; \text{\sffamily{pic$_2$}}, \;\! 
L\hspace{-0.1mm}_{_{\text {\scshape t}}}, \;\!
L\hspace{-0.2mm}_{_{\text {\scshape s}}}, \;\! e \rangle )

 \\
\ \ \ \ \ \ \ \ \ \
\ \ \ \ \ \ \ \ \ \ \ \ \ \ \ \ \ \  \downarrow{ \! \tau} \\
\ \ \ \ (c', \; \langle \repdec, \;\! \textit{do} \;  \text{\sffamily{pic$_1$}} \;\! \textit{until} \; \text{\sffamily{pic$_2$}} , \;\! 
L\hspace{-0.1mm}_{_{\text {\scshape t}}}, \;\!
L\hspace{-0.2mm}_{_{\text {\scshape s}}}, \;\! e \rangle \; ) \vspace{1mm} \\
\text{where} \\
c'= 
[ \hspace{.3mm}
^{\text{{\tiny{ $\perp$}}}}
\hspace{-0.7mm} / \hspace{-0.4mm}
_{\widehat{\widehat{\text{\footnotesize \sffamily{pic}}}}_1.\text{{\scshape \footnotesize src}}}
\hspace{1mm},
\hspace{.1mm}
^{\text{{\tiny{ $\perp$}}}}
\hspace{-0.7mm} / \hspace{-0.4mm}
_{\widehat{\widehat{\text{\footnotesize \sffamily{pic}}}}_1.\text{{\scshape \footnotesize tgt}}}
\hspace{1mm},
\hspace{.1mm}
^{\text{{\tiny{\bf true}}}}
\hspace{-0.7mm} / \hspace{-0.4mm}
_{\text{\footnotesize \sffamily{pic}}_1.\text{{\scshape \footnotesize src}}}
\hspace{.3mm}
 ]
 
}
\end{center}

\end{center}
\end{minipage}
\end{minipage}



\caption{Structured Operational Semantics\label{SOS}}

\end{table}


In \autoref{SOS}, we provide the \sos \hspace{0mm} rules defining a translation from activities
to control graphs. Some comments are in order concerning this table: \\
\vspace{-3mm}\\
~\hspace{-9mm} $\ {\bullet} \ $
the notation for value substitution in control link maps 
requires some explanation:
$c[^\text{{\footnotesize{\bf true}}} / _l]=_{_{def}} \! c'$ where
$c'(l') = c(l)$ for $l \neq l'$ and $c'(l)= \text{{\footnotesize{\bf true}}}$.
By abuse of notation, we also apply value substitution to sets of
control links. Hence, if $\Pi$ is a set of activities,
then, e.g., $c[^\text{{\footnotesize{\bf false}}} / 
\widehat{\Pi}.\text{{\scshape \footnotesize src}}]$
is the substitution whereby any control link
occurring as source of an activity in $\widehat{\Pi}$
has its value set to $\text{{\footnotesize{\bf false}}}$. \\
~\hspace{-8mm} $\! \ \ {\bullet} \ $
the rules for the \text{{\sffamily seq}} activity have
been skipped as 
for any \text{{\sffamily seq}} one can construct a behaviorally equivalent 
\text{{\sffamily flo}} activity having the same set of subactivities. This is mereley achieved by introducing the appropriate control links between any
two consecutive subactivities. \\
~\hspace{-8mm} $ {\bullet} \ $
in the rules for activity \text{\sffamily{flo}} we have dropped the field $\text{{\scshape \footnotesize lnk}}$ 
since its value is constant ($\text{{\scshape \footnotesize lnk}}$ is used to define 
a scope for control link variables). \\
~\hspace{-8mm} $\ {\bullet} \ $
in the rules for the repeat activity \text{{\sffamily rep}}, 
we have introduced a new activity,  \text{{\sffamily unf}} (unfold). 
Its syntax is: $\text{{\sffamily unf}}$  ::= 
$\langle \unfdec, 
do  \ \text{{\sffamily act}} \ then \ 
\text{{\sffamily pic}}_1
\ until \ 
\text{{\sffamily pic}}_2, \  
L\hspace{-0.2mm}_{_{\text {\scshape t}}}, \
L\hspace{-0.2mm}_{_{\text {\scshape s}}}, \ 
e\rangle. \ $
Activity \text{{\sffamily unf}} is introduced as a result of the execution of rule {\scriptsize REP1}. This  mirrors the unfolding of an iteration
by activity $\text{{\sffamily pic}}_1$.  Rule  {\scriptsize UNF1} represents the execution of an action by the unfolded activity
and rule {\scriptsize UNF2} represents the termination of the iteration whereby \text{{\sffamily unf}} is transformed back into a \text{{\sffamily rep}} 
activity identical to the original one.
Note how, in this transformation, all links (strictly) contained in $\text{{\sffamily pic}}_1$ are reset to the undefined value.
The \text{{\sffamily rep}} activity as a whole terminates by means of rule {\scriptsize REP2} representing
the activation of activity $\text{{\sffamily pic}}_2$.\\
~\hspace{-9mm} $\ {\bullet} \ $
as a notational convention, we have used ``$*$'' to denote a wildcard activity, as seen in the rule for dead-path elimination ({\scriptsize DPE}). \\

When applied to the initial control state 
$(c_\text{{\sffamily act}}, \text{{\sffamily act}})$
of a well formed activity $\text{{\sffamily act}}$,
the \sos \hspace{0mm} rules defined in \autoref{SOS} yield a first control graph, the raw control graph, that we note 
$\textbf{r-cg}(\text{{\sffamily act}})$. A transition in this control graph 
will be denoted by 
$(c', \text{ \sffamily{act}}') \xrightarrow[\raisebox{10pt}{\ensuremath{\scriptstyle ^{^\text{ \sffamily{act}}}}}]{ \sigma  }  (c'', \text{ \sffamily{act}}'') $. 

\vspace{-5mm}

\subsection{Properties of Raw Control Graphs}

%
%

\begin{myprop}
The set of states of $\textbf{r-cg}(\text{{\sffamily act}})$ is finite.
\end{myprop}

\vspace{-3.5mm}

\begin{proof}

We can structurally define function $\# \hspace{-.7mm}_{_{ub}}\!(\text{{\sffamily act}})$ that gives an upper
bound for the number of states generated from an activity $\text{{\sffamily act}}$. To obtain this upper bound
we consider an empty set of control links so as to allow the maximum possible interleaving and hence generating the maximum number of states.
This upper bound is structurally defined as follows: \\ 
$\# \hspace{-.7mm}_{_{ub}}\!(\text{{\sffamily nil}})\!\!=\!\!1 \ \ \ \ \ \ \ 
\# \hspace{-.7mm}_{_{ub}}\!(\text{{\sffamily inv}})\!\!=
\!\!\# \hspace{-.7mm}_{_{ub}}\!(\text{{\sffamily rec}})\!\!=
\!\!\# \hspace{-.7mm}_{_{ub}}\!(\text{{\sffamily ses}})\!\!=
\!\!2
\ \ \ \ \ \ \ \ \ 
\#{ \hspace{-.7mm}_{_{ub}}\!}(\langle \repdec,
do \
\text{{\sffamily pic}}_1
\ until \
\text{{\sffamily pic}}_2,
, 
, 
\rangle)\!\!=\! \# \hspace{-.7mm}_{_{ub}}\!(\text{{\sffamily pic}}_1) + \# \hspace{-.7mm}_{_{ub}}\!(\text{{\sffamily pic}}_2) + 1
$ \\
$\# \hspace{-.7mm}_{_{ub}}\!(\langle \flodec,
\text{{\sffamily act}}_1
| \cdots \ |
\text{{\sffamily act}}_n, \
, \
, \
\rangle)=(\# \hspace{-.7mm}_{_{ub}}\!(\text{{\sffamily act}}_1)+1) \times \ldots \times (\# \hspace{-.7mm}_{_{ub}}\!(\text{{\sffamily act}}_n)+1) + 1$\\
$\# \hspace{-.7mm}_{_{ub}}\!(\langle \picdec,
\text{{\sffamily rec}}_1 ; \text{{\sffamily act}}_1
+ \cdots +
\text{{\sffamily rec}}_n ; \text{{\sffamily act}}_n, \
, \
, \
\rangle)
= \# \hspace{-.7mm}_{_{ub}}\!(\text{{\sffamily act}}_1) + 2 + \ldots + \# \hspace{-.7mm}_{_{ub}}\!(\text{{\sffamily act}}_n) + 2 + 1 $\\

\vspace{-5mm}
\end{proof}

\vspace{-2mm}

\begin{myprop}
$\textbf{r-cg}(\text{{\sffamily act}})$ is free of $\tau$ loops. \label{loopfree}
\end{myprop}
%
\begin{proof}
The only case in which a $\tau$ loop could appear is in the $\text{{\sffamily pic}}_1$ of a repeat activity. But since $\text{{\sffamily pic}}$ necessarily 
contains an initial receive activity, any potential $\tau$ loop is broken by this receive activity.
\end{proof}

\begin{myprop}
$\textbf{r-cg}(\text{{\sffamily act}})$ is $\tau$-confluent, \label{confluent} i.e.: \\
$g \xrightarrow[%
\raisebox{10pt}{%
\ensuremath{\scriptstyle ^{^\text{ \sffamily{act}}}}}]
{ \tau  } {\scriptscriptstyle\hspace{-1mm} _{}} \ g_1 \ $ 
and \ 
$g \xrightarrow[%
\raisebox{10pt}{%
\ensuremath{\scriptstyle ^{^\text{ \sffamily{act}}}}}]
{ \sigma  } {\scriptscriptstyle\hspace{-1mm} _{}} \   g_2 \ \ \ \Rightarrow \ \ \ \exists g'  $ 
where 
$g_1 \xrightarrow[%
\raisebox{10pt}{%
\ensuremath{\scriptstyle ^{^\text{ \sffamily{act}}}}}]
{ \sigma  }  {\scriptscriptstyle\hspace{-1mm} _{}} \  g' \ $ 
and \ 
$g_2 \xrightarrow[%
\raisebox{10pt}{%
\ensuremath{\scriptstyle ^{^\text{ \sffamily{act}}}}}]
{ \tau  } {\scriptscriptstyle\hspace{-1mm} _{}} \   g' $

\end{myprop}

\vspace{-8mm}

\begin{proof} \emph{(sketch)} 
Let us consider the situation where from some state 
$(c, \text{\sffamily{act}}_0)$ we can fire two transitions: \\
$(1) \ \ (c, \text{\sffamily{act}}_0) \xrightarrow[%
\raisebox{10pt}{%
\ensuremath{\scriptstyle ^{^\text{ \sffamily{act}}}}}]
{ \tau  }  (c_1, \text{\sffamily{act}}_1) \ \ \ $ 
and \ 
$\ \ (2) \ \ \ (c, \text{\sffamily{act}}_0) \xrightarrow[%
\raisebox{10pt}{%
\ensuremath{\scriptstyle ^{^\text{ \sffamily{act}}}}}]
{ \sigma  }  (c_2, \text{\sffamily{act}}_2). \ \ \ $ \vspace{-6mm}
Since one of the actions is $\tau$ then there is necessarily 
a  subactivity $\text{ \sffamily{flo}} \in \widehat{\text{ \sffamily{act}}_0}$ with  
$\text{ \sffamily{flo}} = \langle \flodec, 
\cdots \ 
\text{{\sffamily act}}'_1 \ 
| \cdots \ | 
\text{{\sffamily act}}'_2 \ 
\cdots, \  
L\hspace{-0.2mm}_{_{\text {\scshape t}}}, \
L\hspace{-0.2mm}_{_{\text {\scshape s}}}, \ 
e
\rangle \ $
where transition $(1)$ is produced by $\text{{\sffamily act}}'_1$
and transition $(2)$ by $\text{{\sffamily act}}'_2$. Since transitions $(1)$ and $(2)$
are not conflicting, then both sequences $(1)$ then $(2)\ \ $ and $\ \ (2)$ then $(1)\ \ $ are 
possible and reach the same state.
\end{proof}

\begin{myprop} ~\\
(i) All sink states of \textbf{r-cg}(\text{{\sffamily act}}) (i.e. states with no outgoing transitions) are of the form (c, \mynil). 
(Hence sink states may differ only by their control link maps),\\ 
(ii) for any state of \textbf{r-cg}(\text{{\sffamily act}}), there exists a path that leads to a sink state.
\end{myprop}

\begin{proof} \emph{(sketch)}\\
$(i)$ Let us consider the non trivial case where $\text{{\sffamily act}}$ is not an atomic activity, and
a state $(c, \text{{\sffamily act}}_0)$ reachable from the initial state 
$(c_\text{{\sffamily act}}, \text{{\sffamily act}})$.
If $\text{{\sffamily act}}_0.\text{{\scshape \footnotesize beh}} = \mynil$
then $(c, \text{{\sffamily act}}_0)$ cannot be a sink state since it can make a transition to $(c, \mynil)$.
If $\text{{\sffamily act}}_0.\text{{\scshape \footnotesize beh}} \neq \mynil$ then
surely there is one activity  $\text{{\sffamily act}}' \in \text{{\sffamily act}}_0.\text{{\scshape \footnotesize beh}}$
which is \emph{head} and ready for execution in the context of control link map $c$. This stems from
the non cyclicity property which, it can be proven, is preserved along the execution path from 
the initial state. This means that all the activities in $\widehat{\text{\sffamily act}}$ that 
preceed $\text{{\sffamily act}}'$ have been either executed or cancelled.
Hence, $(c, \text{{\sffamily act}}_0)$ can make a transition (the one derived from  $\text{{\sffamily act}}'$),
and thus cannot be a sink state. \\
$(ii)$ the inspection of all \sos \hspace{0mm} rules shows that for all states
$(c, \text{{\sffamily act}}_0)$, where $\text{{\sffamily act}}_0$ is not a repeat activity,
there is a transition to some other state
$(c', \text{{\sffamily act}}')$ where $\text{{\sffamily act}}'$ is strictly (syntactically)
simpler (i.e., smaller in size) that $\text{{\sffamily act}}_0$. In the case of repeat,  there is also
always a path leading to a syntactically simpler activity, which is through 
the pre-empting activity $\text{{\sffamily pic}}_2$. $\mynil$ being the simplest
activity, thus all activities must reduce to  $\mynil$.
%

\end{proof}

\vspace{-8mm}

\begin{figure}[!ht]

\begin{center}
\includegraphics[scale=0.65]{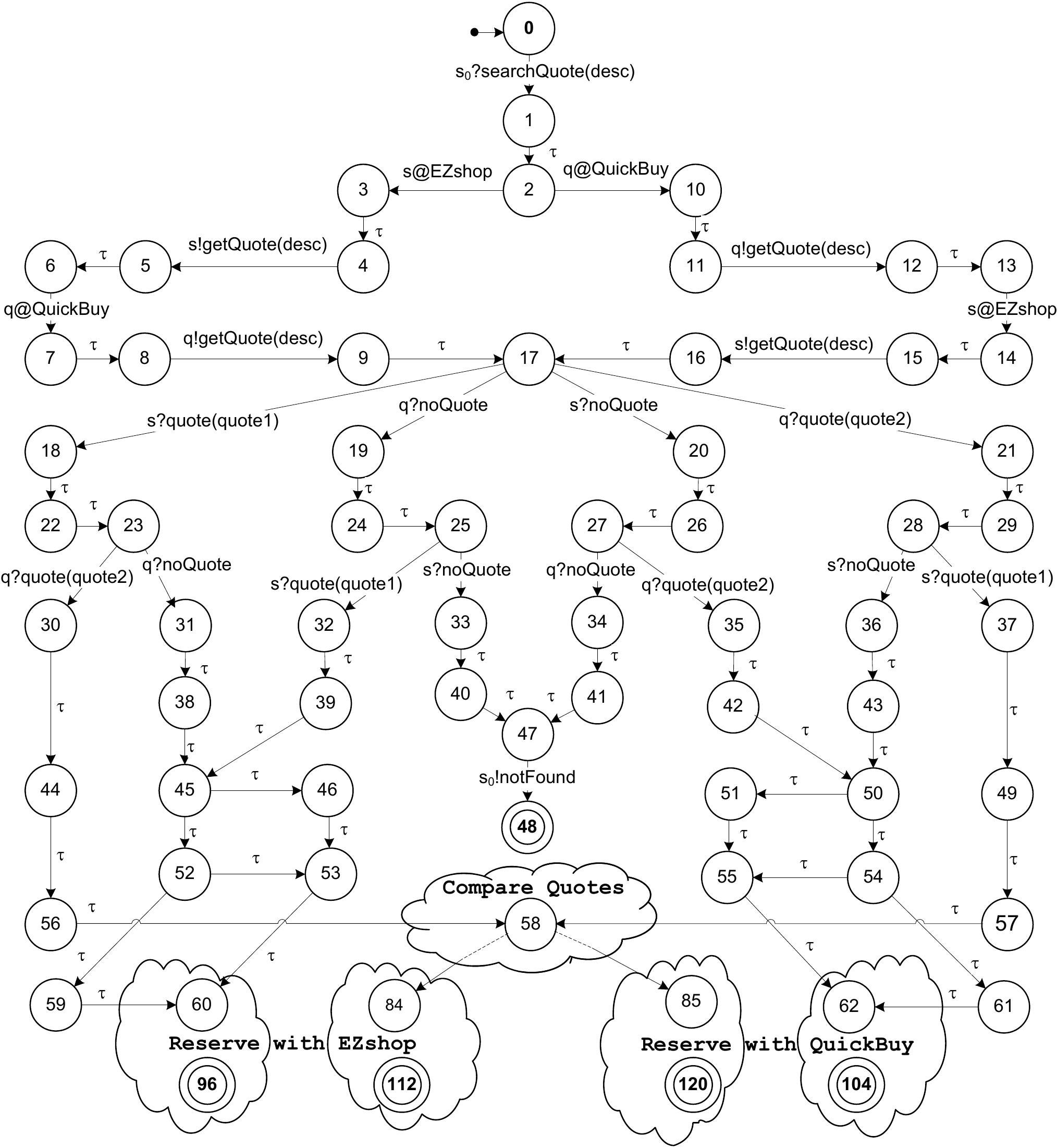}
\vspace{-5mm}

\caption{The QuoteComparer Control Graph\label{quotecomparergraph}}

\end{center}

\end{figure}

\vspace{-2mm}

\subsection{Control Graph Transformations}

The first transformation applied to $\textbf{r-cg}(\text{{\sffamily act}})$
is $\tau$-prioritization and results in control graph $\tau\hspace{-0.1mm}\textbf{p-cg}(\text{{\sffamily act}})$.
Then we apply $\tau$-compression, resulting in control graph $\tau\hspace{-0.1mm} \textbf{c-cg}(\text{{\sffamily act}})$ that is free of $\tau$ transitions.
Next comes a transformation into run-to-completion semantics  resulting in 
$\textbf{rtc-cg}(\text{{\sffamily act}})$. Finally, strong equivalence
minimization is applied resulting in $\textbf{cg}(\text{{\sffamily act}})$, a graph with a single sink state.

\vspace{-4mm}

\subsubsection{$\tau$-Priorisation and $\tau$-Compression}

In~\cite{springerlink:10.1007/3-540-44612-5_34}, the authors give an efficient algorithm that, given a graph in which the $\tau$ transitions are confluent and in which there are no $\tau$ loops, prioritises $\tau$ transitions in the graph while preserving branching equivalence. Given Property \autoref{confluent}, i.e that the $\tau$ transitions of any  $\textbf{r-cg}(\text{{\sffamily act}})$ are confluent, and Property \autoref{loopfree}, i.e there are no $\tau$ loops in any $\textbf{r-cg}(\text{{\sffamily act}})$, we can apply the algorithm given in~\cite{springerlink:10.1007/3-540-44612-5_34} and obtain a $\tau$-prioritised control graph that we call $\tau\hspace{-0.1mm}\textbf{p-cg}(\text{{\sffamily act}})$. In such a graph, the outgoing transitions from a state are either all $\tau$ transitions, or they are all observable actions. In~\cite{springerlink:10.1007/3-540-44612-5_34}, an algorithm for $\tau$-compression is also given, essentially leaving the graph $\tau$ transition free. We can therefore apply $\tau$-compression to $\tau\hspace{-0.1mm}\textbf{p-cg}(\text{{\sffamily act}})$ and obtain a $\tau$-free 
branching equivalent control graph that we call $\tau\hspace{-0.1mm} \textbf{c-cg}(\text{{\sffamily act}})$. 

\vspace{-4mm}

\subsubsection{Applying run-to-completion semantics}

Here we process control graphs so as to verify the \textit{run-to-completion} property. This property implies the following behavior of control graphs: 
after a receive, all possible invocations, session initiations, or $\tau$ actions are executed before another message can be received. 

In order to give \seb \hspace{0mm} a run-to-completion semantics, we give a priority order to the transitions in a $\tau$-prioritised control graph $\tau\hspace{-0.1mm}\textbf{p-cg}(\text{{\sffamily act}})$ (this also applies for $\tau$-compressed control graphs). We consider the outgoing transitions from a state and we define the following priority order between the actions that label these transitions:
s!op(x) $>$ s@p $>$ s?op(x). This means that when two transitions labelled with actions with different priorities
are possible from a given state, then the one having
a lower priority is pruned. All states that have become unreachable from the initial state are also pruned.  \\ 

\label{controlgraphsofsebactivities}

\autoref{quotecomparergraph} shows the control graph generated for the QuoteComparer \seb \hspace{0mm} service shown in \autoref{quotecomparerdiag} 
in which the $\tau$-priorisation and run-to-completion semantics transformations have been applied. 
$\tau$-compression was not applied here for the sake of illustration.
In this figure, the computation between states 0 and 17 represents an example of run-to-completion.

\begin{figure}[!ht]

\begin{center}

\includegraphics[scale=0.7]{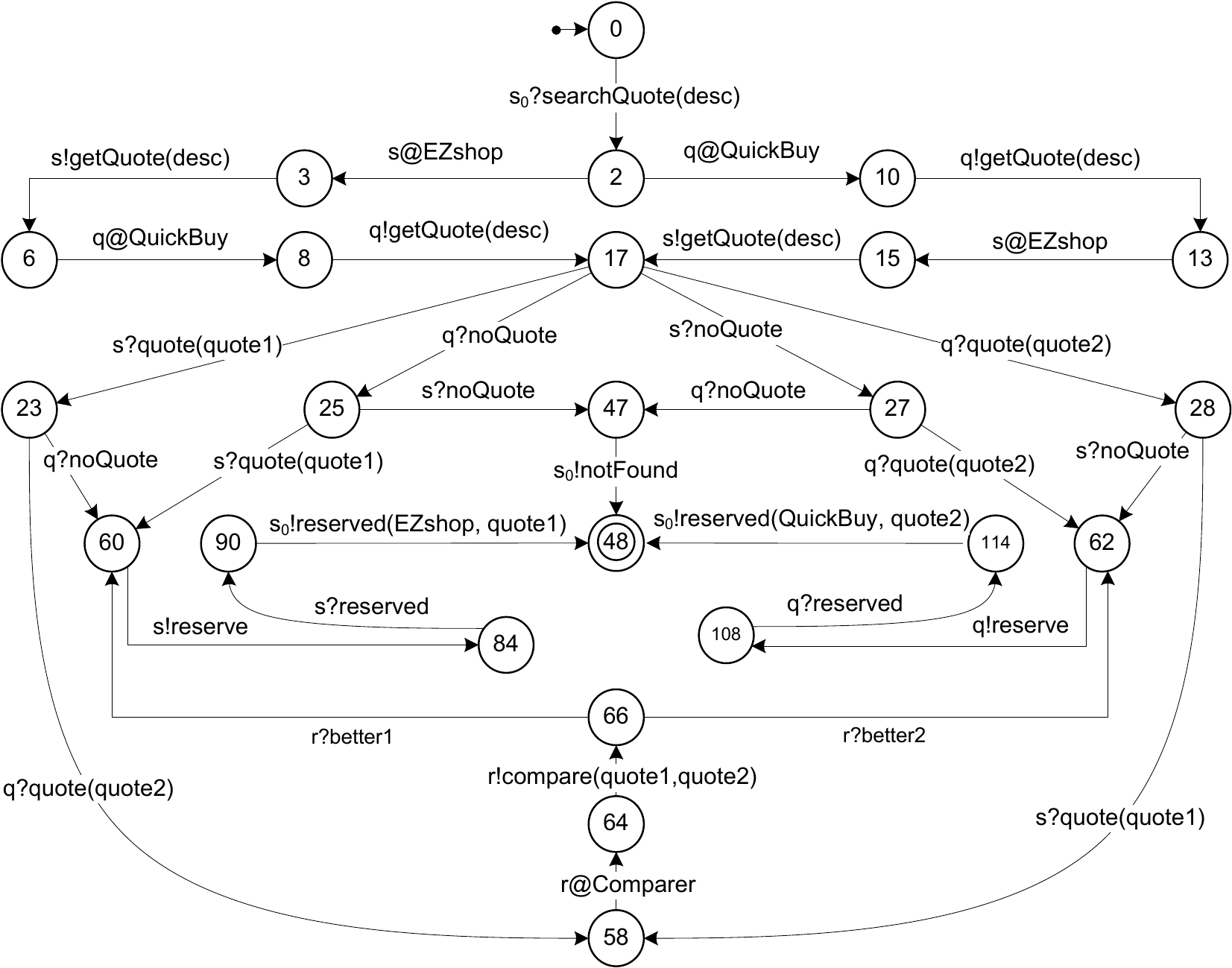}
\caption{The QuoteComparer Reduced Control Graph\label{quotecomparergraphreduced}}
\end{center}
\end{figure}

Note that the control graph of \autoref{quotecomparergraph} contains 5 terminal states. 
They all correspond to couples of the form $(c_i, \mynil)$, in accordance with Property 4. They differ only by their control 
link maps. For example, state 48 corresponds to the couple $(c_{48}, \mynil)$ where 
$c_{48}$ is given by
$c_{48}(l_6)=c_{48}(l_7)= c_{48}(l_4) = c_{48}(l_5) = true$ and where 
$c_{48}$ is false for the remaining links.

\vspace{-4mm}

\subsubsection{Minimisation}

When strong (or branching)  equivalence minimisation is applied to control graph
$\textbf{rtc-cg}(\text{{\sffamily act}})$, a minimal control graph $\textbf{cg}(\text{{\sffamily act}})$ is produced that has one and only sink state (because all sink states are equivalent). \autoref{quotecomparergraphreduced} shows the minimised control graph of the QuoteComparer example. We used the CADP~\cite{springerlink:10.1007/3-540-61474-5_97} toolset to perform some of the transformations defined in this section on the QuoteComparer example. \\ 

\vspace{-4.5mm}

\subsection{Semantics of activities}

Henceforth, when we write $\textbf{cg}(\text{{\sffamily act}})$ for any well-formed activity {\sffamily act}, we 
will consider that we are dealing with the $\tau$-compressed, run-to-completion and minimised control graph,
and we name its unique sink state 
$\textbf{term}(\text{{\sffamily act}})$, to be referred to as the terminal state. We also adopt the following notations: ${\bf init}(\text{\sffamily{act}})$
denotes the initial state of ${\bf cg}(\text{\sffamily{act}})$;
${\bf states}(\text{\sffamily{act}})$ is the set of  states of 
${\bf cg}(\text{\sffamily{act}})$; ${\bf trans}(\text{\sffamily{act}})$
is the set of transitions of ${\bf cg}(\text{\sffamily{act}})$.
A transition in this final control graph $\textbf{cg}(\text{{\sffamily act}})$ 
will be denoted by $g \xrightarrow[\raisebox{10pt}{\ensuremath{\scriptstyle ^{^\text{ \sffamily{act}}}}}]{ \sigma  }  g' $. 

\begin{mydef} \emph{Open for reception} - A state $g$ of ${\bf cg}(\text{\sffamily{act}})$ is said to be 
open for reception on session $s$ and we note 
$open(\text{\sffamily{act}}, g, s)$, 
$if\!f$ state $g$ has at least one outgoing transition 
labeled with a receive action on session $s$. More formally: 
$open(\text{\sffamily{act}}, g, s) =_{def}$ $\exists$ $op, \ x_1 \ \ldots \ x_n, \ g'$ 
such that $g \xrightarrow[%
\raisebox{10pt}{%
\ensuremath{\scriptstyle ^{^\text{ \sffamily{act}}}}}]
{ s?op(x_1, \cdots, x_n)  }  g' $. 
\end{mydef}

\vspace{-4mm}

\subsection{Free, bound, usage and forbidden occurrences of variables}


\begin{mydef} \emph{Variables of an activity} - For an activity $\text{\sffamily{act}}$ we define 
the set of variables occurring in $ \text{\sffamily{act}}$: 
$\ \ V(\text{\sffamily{act}})=_{\text{\tiny def}} 
\{ \ z \ | \ z \ \text{occurs in} \ \text{\sffamily{act}} \}. $
\end{mydef}

\begin{mydef} \emph{Binding occurrences} - For variables 
$y \in V(\text{\sffamily{act}})$,
$s \in V(\text{\sffamily{act}})$ and
$p \in V(\text{\sffamily{act}})$, 
the following underlined occurrences
are said to be binding occurrences in $\text{\sffamily{act}}$:
$\underline{s}@p$, $s?op(\cdots, \underline{y}, \cdots)$ and
$s?op(\cdots, \underline{p}, \cdots)$.
We denote $BV(\text{\sffamily{act}})$ the set
of variables having a binding occurrence in 
$\text{\sffamily{act}}$. 
\end{mydef}

\begin{mydef} \emph{Usage occurrences} - For variables 
$y \in V(\text{\sffamily{act}})$,
$s \in V(\text{\sffamily{act}})$ and
$p \in V(\text{\sffamily{act}})$, 
the following underlined occurrences
are said to be usage occurrences in $\text{\sffamily{act}}$:
$s@\underline{p}$, $\underline{s}?op(\cdots)$,
$\underline{s}!op(\cdots)$, 
$s!op(\cdots, \underline{p}, \cdots )$ and $s!op(\cdots, \underline{y}, \cdots )$.
We denote $UV(\text{\sffamily{act}})$ the set
of variables having at least one usage occurrence in 
$\text{\sffamily{act}}$.\\
\end{mydef}

\vspace{-6mm}

\begin{mydef} \emph{Free occurrences} - A variable 
$z \in V(\text{\sffamily{act}})$
is said to occur free in $\text{\sffamily{act}}$,
$if\!f$ there is a path in ${\bf cg}(\text{\sffamily{act}})$: 
$init(act) \xrightarrow[%
\raisebox{-1pt}{%
\ensuremath{\scriptscriptstyle 
^{^\text{ \sffamily{act}}}}}]{\ \sigma_1\ } g_1,
\cdots, g_{n-1} \xrightarrow[%
\raisebox{-1pt}{%
\ensuremath{\scriptscriptstyle 
^{^\text{ \sffamily{act}}}}}]{\ \sigma_n\ } g_n$
where $z$ has a usage occurrence in $\sigma_n$
and has no binding occurrence in any 
of $\sigma_1, \cdots, \sigma_{n-1}$.
We denote $FV(\text{\sffamily{act}})$ the set
of variables having at least one free occurrence in 
$\text{\sffamily{act}}$. 
\end{mydef}

\begin{mydef} \emph{Forbidden occurrences} - $op?(\cdots p_0 \cdots)$ and $s_0@p$
are forbidden occurrences. As we shall explain later, $p_0$ is reserved
for the own location of the service, while $s_0$ is a reserved session
variable that receives a session id implicitly at service instantiation
time.
\end{mydef}

\vspace{-4mm}


\section{Syntax and Semantics of Service Configurations}

\seb\hspace{0mm} activities can become deployable
services that can be part of configurations of services. These configurations have a dynamic semantics, based on which we can define the property of safe interaction.

Let $M$ be a partial map from variables 
$V\!ar$ to  $V\!al \ \cup \{ \perp \}$, the set of values augmented
with the undefined value. Henceforth, we consider couples 
$(M, \hspace{0.3mm} \text{\sffamily{act}})$
where $dom(M)=V\!(\text{\sffamily{act}})$.

\vspace{-1mm}

\subsection{Deployable Services}
\vspace{-1mm}


\begin{mydef} \emph{Deployable services} - The couple 
$\lb M, \hspace{0.3mm} \text{\sffamily{pic}}\rb$
is a deployable service iff: \vspace{2mm} \\
\begin{tabular}{lll}
\noindent
${\scriptstyle \bullet}$ 
$p_0 \in FV\!(\text{\sffamily{pic}}), \ s_0 \in FV\!(\text{\sffamily{pic}})$ & 
${\scriptstyle \bullet}$ 
$FV\!(\text{\sffamily{pic}}) \cap SesV\!ar \!=\! \{ s_0 \} \ \ $ & 
${\scriptstyle \bullet}$
$\text{\sffamily{pic}}.\text{{\scshape beh}}= 
\sum \! s_0?op_i(\tilde{x_i}); \text{\sffamily{act}}_i \ \ $ \vspace{1mm} \\ 
${\scriptstyle \bullet}$
$dom(M)=V\!(\text{\sffamily{pic}})$ & 
${\scriptstyle \bullet}$
$\forall z \in V\!(\text{\sffamily{pic}}) {\scriptstyle \ \setminus \ } FV\!(\text{\sffamily{pic}}): M(z)= \perp \ \ $ & 
${\scriptstyle \bullet}$
$\forall z \! \in \! FV\!(\text{\sffamily{pic}}) 
{\scriptstyle \ \setminus \ }  \{ s_0 \}: 
m(z) \! \neq \perp \ \ $ 
\end{tabular}
\end{mydef}

\begin{quote}
\textit{Informally, $(M,\text{{\sffamily act}})$ is a deployable service if $\text{{\sffamily act}}$ is a $\text{{\sffamily pic}}$, $s_0$ is its 
only free session variable, the initial receptions of $\text{{\sffamily pic}}$ are on session $s_0$, its location is defined (variable $p_0$ is set in $M$\!), and all its free variables, except $s_0$, 
have defined values in $M$.}
\end{quote}

\vspace{-1mm}

\begin{mydef} \emph{Running service instances} - The running state of a service instance derived from the deployable service  
$\lb M, \hspace{0.3mm} \text{\sffamily{pic}} \rb$
is the triple
$(m, \hspace{0.3mm} \text{\sffamily{pic}}
\hspace{0.7mm}
_{\scriptscriptstyle ^\blacktriangleright} g
)$
where: $g \in \text{\bf states}(\text{\sffamily{pic}})$ \; and \; $dom(m)=dom(M)$. The initial state of this running service instance is the triple $(M[ \hspace{0.2mm}
^\beta \hspace{-1mm}
/
\hspace{-0.7mm} _{s_0}
\hspace{-0.1mm}
], \hspace{0.3mm} \text{\sffamily{pic}}
\hspace{0.7mm}
_{\scriptscriptstyle ^\blacktriangleright} \text{\bf init}(\text{\sffamily{pic}})
)$ with $\beta \ fresh$. A deployed service behaves like a factory creating a new running service instance each time it receives a session initiation request.
\end{mydef}

\subsection{Service Configurations}
\vspace{-1mm}

Configurations are built from deployable services. We provide an abstraction of the communication bus that is necessary to formalize and prove the desirable properties of service configurations. Service instances exchange messages
through FIFO queues, and session bindings
are set up in order to establish the corresponding queues.

\begin{mydef} \emph{Service configurations} - When deployed, a set of deployable services yields a configuration noted: 
$\lb M_1, \text{\sffamily{pic}}_1 \rb \ \diamond \
\cdots \ \diamond \  
\lb M_k, \text{\sffamily{pic}}_k \rb$. The symbol $\diamond$ denotes the associative and commutative deployment operator, meaning that services are deployed together and share the same address space.
\end{mydef}

\begin{mydef} \emph{Well-partnered service configuration} - A service configuration $ 
\lb M_1, \text{\sffamily{pic}}_1 \rb \ \diamond \
\cdots \ \diamond \  
\lb M_k, \text{\sffamily{pic}}_k \rb$ 
is said to be well partnered 
iff: \\
\begin{tabular}{cc} 
${\scriptstyle \bullet}$
$\forall i, j: i \neq j 
\Rightarrow M_i(p_0) \neq M_j(p_0)$ & 
${\scriptstyle \bullet}$
$\forall i, p: M_i(p) \neq \perp \Rightarrow \exists j 
\text{ with } M_i(p) = M_j(p_0)$ 
\end{tabular}
\end{mydef}

\begin{quote}
\textit{That is, any two services have different location addresses, and any partner required by one service is present in the set of services.}
\end{quote}

\subsection{Running Configurations}
\vspace{-1mm}

\begin{mydef} \emph{Message queues} - ${\cal Q}$ is a set made of message queues with  \ 
${\cal Q}$ ::= $q$ $|$ $q \diamond {\cal Q}$, where $q$ is an individual 
{\small FIFO} message queue of the form 
$q$ ::= $\delta \hookleftarrow \widetilde{M\!e\!s}$ with 
$\widetilde{M\!e\!s}$ a possibly empty list of ordered messages
and $\delta$ the destination of the messages in the queue. 
The contents of $\widetilde{M\!e\!s}$ depend on the kind
of the destination $\delta$.
If $\delta$ is a service location, $\widetilde{M\!e\!s}$ 
contains only session initiation requests of the form $new(\alpha)$. 
\noindent
However if $\delta$ is a session id, 
then $\widetilde{M\!e\!s}$  contains only operation messages of the form 
$op(\tilde{w})$. 
\end{mydef}

\begin{mydef} \emph{Session bindings} - A session binding is an unordered pair of session ids 
($\alpha$,~$\beta$). 
A running set of session bindings
is noted ${\cal B}$ and has the syntax
${\cal B}$ ::= $(\alpha,~\beta)~|~(\alpha, \beta)~\diamond~{\cal B}$. 
If $(\alpha, \beta)
\in {\cal B}$ then
$\alpha$ and $\beta$ are said to be bound
and messages sent on local session id $\alpha$ are 
routed to a partner holding local session id $\beta$, and vice-versa. 
\end{mydef}

\begin{mydef} \emph{Running configurations} - A running configuration, $C$, is a configuration made of services, service
instances, queues and bindings all 
running concurrently and sharing the same address space: \\ 
${\cal C} =  {\cal C}_{serv}  \diamond (m_1, \text{\sffamily{act}}_1, g_1) {\scriptstyle \cdots} (m_k, \text{\sffamily{act}}_k, g_k) \diamond {\cal Q} \diamond {\cal B}$ 
where ${\cal C}_{serv} = \lb M_1, \text{\sffamily{pic}}_1 \rb \cdots \lb M_n, \text{\sffamily{pic}}_n \rb $ is a well-partnered service configuration, and $(m_1, \text{\sffamily{act}}_1, g_1) {\scriptstyle \cdots} (m_k, \text{\sffamily{act}}_k, g_k)$ are service instances.
\end{mydef}

Again, operator $\diamond$ is associative and commutative hence the order of services, instances, bindings and queues is irrelevant. Also if the sets of bindings or queues are empty, they are omitted. 

\begin{mydef} \emph{Initial Running configuration} - A service configuration cannot bootstrap itself as this requires at least one client instance that begins by opening a session. $(m, \text{\sffamily{act}}, g)$ is one such client where $\text{\sffamily{act}}.\text{{\footnotesize\scshape beh}} = s@p; \text{\sffamily{act}}'$ and $ \exists i$ such that $ m(p)=M_i(p_0) $. Hence the minimal initial running configuration is: 
$\ \ C = \lb M_1, \text{\sffamily{pic}}_1 \rb \cdots \lb M_n, \text{\sffamily{pic}}_n \rb \diamond (m, \text{\sffamily{act}}, g).$
\end{mydef}



\subsection{Semantics of Service Configurations}

Here we provide a full semantics for \seb \hspace{0mm} that allows us to formalize the property that we want to assess in \seb \hspace{0mm} programs.
The service configuration semantics is defined using the  four \sos \hspace{0mm} rules in \autoref{rulesnetworkedservices}.

\begin{table}
\fontsize{10pt}{9pt}
\selectfont

\begin{center}
\regle{
\vspace{-2mm} \\
\hspace{-5mm}
\fbox{{\tiny SES1}}
\vspace{-4.5mm}
\ \ \ \ \ \ \ \ \ \ \ \ \ \
g \xrightarrow[%
\raisebox{10pt}{%
\ensuremath{\scriptscriptstyle ^{^\text{ \sffamily{act}}}}}]{\ s@p \ } g'
\ \ \ \

\ \ \ \alpha, \beta \  \text{fresh}
}
{
\vspace{-1.5mm} \\
\ \ \ \ \ \ 
{\scriptstyle \cdots} \
( m, \text{\sffamily{act}} \
 _{\scriptscriptstyle ^\blacktriangleright}   g) \
{\scriptstyle \cdots} \ 
 {\scriptstyle \; \diamond}
\ 
{\scriptstyle \cdots} 
\ 
 (m(p) \hookleftarrow
{\scriptstyle \widetilde{M\!e\!s}}
) \ {\scriptstyle \cdots} \
 {\scriptstyle \; \diamond}
 \ {\cal B} \
 \ \   \longrightarrow  \vspace{2mm} \\
 {\scriptstyle \cdots} \
( m
[ \hspace{0.2mm}
^\alpha \hspace{-0.9mm}
/
\hspace{-0.6mm} _s
\hspace{-0.1mm}
],
\hspace{0.3mm}
\text{\sffamily{act}} \
_{\scriptscriptstyle ^\blacktriangleright} g' )
 \ {\scriptstyle \cdots}  \
 {\scriptstyle \; \diamond}
\ 
{\scriptstyle \cdots} (m(p) \hookleftarrow
{\scriptstyle \widetilde{M\!e\!s}}
 \! \cdot \!new(\beta) \; ) \  {\scriptstyle \cdots} \
 {\scriptstyle \; \diamond}
\ {\cal B}  {\scriptstyle \; \diamond} \; (\alpha, \beta)  \
   \\

}{}
\end{center}


\begin{center}
\regle{
\vspace{-1mm} \\
\fbox{{\tiny SES2}}
\vspace{1.1mm}
\ \ \ \ \ \ \ \ \ \  \ \ \ \ \ \ \ \ \ \ \  \ \ \ \ \ \ 
M(p_0) = \pi \ \ \ \ \  \
\ \ \ \
\ \ \ \ \ \ \ \ \ \ \ \ \ \ \ \ \ \ \ \
}
{
\vspace{-1.5mm} \\

\ {\scriptstyle \cdots} \ \lb M, \hspace{0.3mm} \text{\sffamily{pic}} \rb \
{\scriptstyle \cdots} \

 {\scriptstyle \; \diamond}
\ 
\ {\scriptstyle \cdots} \ (\pi \hookleftarrow new(\beta)\!
\cdot \! {\scriptstyle \widetilde{M\!e\!s}}) \ {\scriptstyle \cdots} \

 {\scriptstyle \; \diamond} \ {\cal B} \

\ \ \ \longrightarrow \ \ \  \vspace{2mm} \\
\

\ 
{\scriptstyle \cdots} \ 
\lb M, \hspace{0.3mm} \text{\sffamily{pic}} \rb
\ 
{\scriptstyle \diamond}
\ 
(M
[ \hspace{0.2mm}
^\beta \hspace{-1mm}
/
\hspace{-0.7mm} _{s_0}
\hspace{-0.1mm}
],
\hspace{0.3mm}
\text{\sffamily{pic}} \
_{\scriptscriptstyle ^\blacktriangleright} \hspace{0.2mm}
\text{init}(\text{\sffamily{pic}}) \ )
\ 
{\scriptstyle \cdots}
\ 
 {\scriptstyle \; \diamond}
\ 
{\scriptstyle \cdots} 
\ 
(\pi \! \hookleftarrow \!
{\scriptstyle \widetilde{M\!e\!s}}) \ {\scriptstyle \cdots} \ 

 {\scriptstyle \; \diamond}

\ {\cal B} \

\\
       
}

\end{center}


\begin{center}
\regle{
\vspace{-2mm} \\
\fbox{{\tiny INV}} \ \ \ \ \ \ \ \ \ \
g \xrightarrow[%
\raisebox{10pt}{%
\ensuremath{\scriptscriptstyle ^{^\text{ \sffamily{act}}}}}]{\
s!op(x_1, {\scriptscriptstyle \cdots}, \hspace{0.3mm} x_n) \ }
g'
\ \ \ \
 \ \ \
\ \ (m(s), \beta) \in {\cal B}
\vspace{-4.5mm}
}
{
\vspace{-1.5mm} \\
\ \ \ \ \ \ 
{\scriptstyle \cdots} \
( m, \text{\sffamily{act}} \
 _{\scriptscriptstyle ^\blacktriangleright}   g) \
{\scriptstyle \cdots}  \
 {\scriptstyle \; \diamond}
\ 
{\scriptstyle \cdots}  
\
(\beta \hookleftarrow
{\scriptstyle \widetilde{M\!e\!s}}
) {\scriptstyle \cdots} \
 {\scriptstyle \; \diamond}
 \ {\cal B} \
  \ \ \  \longrightarrow  \vspace{2mm} \\
 \ {\scriptstyle \cdots} \
( m,
\hspace{0.3mm}
\text{\sffamily{act}} \
_{\scriptscriptstyle ^\blacktriangleright} g' )
 \ {\scriptstyle \cdots}  \
 {\scriptstyle \; \diamond}
\ 
{\scriptstyle \cdots} \ (\ \beta \hookleftarrow
{\scriptstyle \widetilde{M\!e\!s}}
 \! \cdot \!
 {\scriptstyle op (  m(x_1), \; \cdots, \; m(x_n) )}  
\; ) \; {\scriptstyle \cdots} \
 {\scriptstyle \; \diamond}
\ {\cal B} \
   \\
       
}
\end{center}


\begin{center}
\regle{
\vspace{-2mm} \\
\fbox{{\tiny REC}} \ \ \ \ \ \ \ \ \ \
\ \ \ \
g \xrightarrow[%
\raisebox{10pt}{%
\ensuremath{\scriptscriptstyle ^{^\text{ \sffamily{act}}}}}]{\
s?op(x_1, {\scriptscriptstyle \cdots}, \hspace{0.3mm} x_n)
\ }
g'
\ \ \ \ \ \ \
\ \ \ m(s)  =  \beta
\vspace{-4.5mm}
}
{
\vspace{-1.5mm} \\
{\scriptstyle \cdots} \
( m, \text{\sffamily{act}} \
_{\scriptscriptstyle ^\blacktriangleright}   g) \ 
{\scriptstyle \cdots}
\ 
 {\scriptstyle \; \diamond}
\ 
{\scriptstyle \cdots}  
\ 
(\beta \hookleftarrow
op(w_1, {\scriptscriptstyle \cdots}, \hspace{0.3mm} w_n)
{\scriptstyle  \cdot \widetilde{M\!e\!s}}
) 
\ 
{\scriptstyle \cdots} 
\ 
 {\scriptstyle \; \diamond}
 \ {\cal B} \
  \ \ \  \longrightarrow  \vspace{2mm} \\
 \ {\scriptstyle \cdots} \
( m
[ \hspace{0.2mm}
^{w_1} \hspace{-0.9mm}
/
\hspace{-0.6mm} _{x_1}
\hspace{-0.1mm},
{\scriptscriptstyle \cdots},
 \hspace{0.2mm}
^{w_n} \hspace{-0.9mm}
/
\hspace{-0.6mm} _{x_n}
\hspace{-0.1mm}
],
\hspace{0.3mm}
\text{\sffamily{act}} \
_{\scriptscriptstyle ^\blacktriangleright} g' )
 \ {\scriptstyle \cdots}  \
 {\scriptstyle \; \diamond}
\ 
{\scriptstyle \cdots} 
\ 
(\beta \hookleftarrow
{\scriptstyle \widetilde{M\!e\!s}}
) \; {\scriptstyle \cdots} \
 {\scriptstyle \; \diamond}
\ {\cal B} \
   \\
       
}
\end{center}

\caption{\sos \hspace{0mm} Rules for Service Configurations}
\label{rulesnetworkedservices}

\end{table}

\vspace{1.3mm}

{\footnotesize SES1} applies when a service instance has a session initiation 
transition, $s@p$, where $m(p)$ is the address of the remote service. The result is that the two fresh session ids $\alpha$ and $\beta$ are created for the local and distant session ends, and message $new(\beta)$ is added to the tail of the message queue targeting service $\pi$. 
{\footnotesize SES2} applies when message $new(\beta)$ is at the head of
the input queue of the service located at $\pi$ (for which $m(p_0)=\pi)$.
The result is that the message is consumed and a new service instance
is created with root session $s_0$ set to $\beta$.
{\footnotesize INV} states that when a service
instance is ready to send an invocation message over session $s$, then the message is appended to the queue whose target
is $\beta$ which is bound to $m(s)$. {\footnotesize REC}
is symmetrical to {\footnotesize INV}.

\subsection{Interaction-safe Service Configurations}

The property of \textit{interaction safety} is verified when no service instance ever reaches a state in which it is awaiting a message, but the message at the head of the corresponding input queue is not expected. The definition that follows is given in two parts, the second depends on the first. 

\begin{mydef}
\emph{One-step Interaction-safe Running Configurations} - A running configuration \\ $~ \ \ \ \ \ {\cal C} = \lb {M}_1, \text{\sffamily{pic}}_1 \rb {\scriptsize \cdots} \lb {M}_n, \text{\sffamily{pic}}_n \rb \diamond ({m}_1, \text{\sffamily{act}}_1  \
_{\scriptscriptstyle ^\blacktriangleright} g_1) {\scriptsize \cdots} ({m}_k, \text{\sffamily{act}}_k \ _{\scriptscriptstyle ^\blacktriangleright} g_k) \diamond {\cal Q} \diamond {\cal B} $ \\ 
is one-step interaction-safe iff for any (service or client) instance $j, \ \ $ 
any session variable $s$, and any operation $op$ the following implication holds: \\
~\\
\begin{tabular}{lcr}
$m_j(s) \hookleftarrow op(w_1, {\scriptstyle \ldots}, w_l) 
\cdot \widetilde{M\!e\!s} \ \  \in \ \ {\cal Q} \ \ $ 
(for some values $w_1,\ldots,w_l$), & and &
$open(\text{\sffamily{act}}_j,g_j,s)$
%
\end{tabular}
\\ $\Rightarrow$  $ \ \ \ $
$g_j \xrightarrow[%
\raisebox{10pt}{
\ensuremath{\scriptstyle ^{^\text{act$_j$}}}}] { s?op(x_1',\ldots,x_l')  } \ \ \  $ (for some variables $x'_1,\ldots,x'_l$)
\end{mydef}


\begin{mydef} \emph{Interaction-safe Service Configuration} - A service configuration $C\!_{ser\!v}$ is interaction-safe iff for any client instance $(m, \text{\sffamily{act}}, g)$ and any running configuration $C$ reachable from $C\!_{ser\!v} \diamond (m, \text{\sffamily{act}}, g)$, $C$ is one-step interaction-safe.
\end{mydef}

\vspace{-5mm}


\section{Related work}
\vspace{-3mm}


The potential of sessions in programming languages has gained recognition recently, and languages such as Java \cite{springerlink:10.1007/978-3-642-21464-6_8} and Erlang \cite{springerlink:10.1007/978-3-642-21464-6_7} have been extended in order to support them. In the service orchestration community, a significant body of work looks at formal models that support sessions for services as a first-class element of the language, such as in the Service-Centered Calculus (SCC) \cite{SCC}, CaSPiS \cite{springerlink:10.1007/978-3-540-68863-1_3}, and Orcharts \cite{coordination}, among others. To the best of our knowledge, our work on \seb \hspace{0mm} is the first that has introduced sessions into the widely adopted \bpel \hspace{0mm} \cite{std/ws-bpel2} orchestration language.


Other formalizations of \bpel \hspace{0mm} have been suggested. For instance, in \cite{DBLP:conf/IEEEscc/KoppKL08} the authors defined an algorithm to derive data-links from \bpel \hspace{0mm} code by evaluating the control flow of processes as described by their control links. In our work, we have taken a step further and given an overall static semantics of variables in \bpel \hspace{0mm}, which has allowed us to define the notion of a well structured activity. \bpel \hspace{0mm} models are quite often based on Petri nets \cite{DBLP:journals/scp/OuyangVABDH07, FASE2005}, which lend themselves well to the task of formalizing control links. By separating our semantics into two steps, we were able to propose a formal specification of \bpel \hspace{0mm} with control links more concisely than with Petri nets, notably when it comes to 
capturing the behavior of dead-path elimination. Other work suggests a formalization that takes into account typing of \bpel \hspace{0mm} process with \wsdl \hspace{0mm} descriptors \cite{DBLP:journals/fmsd/LapadulaPT11}. The session types defined in the present paper can model \wsdl \hspace{0mm} descriptors, and they add the possibility to model a service's behavior.

	
\bpel \hspace{0mm} \cite{std/ws-bpel2} uses correlation sets rather than sessions to relate messages belonging to a particular instance of an interaction. Messages that are logically related are identified as sharing the same \emph{correlation data} \cite{COWS}. We could have defined a ``session oriented'' style for \bpel\hspace{0mm} implemented with correlation sets, but this approach would not lend itself easily to behavioural typing which is our objective for \seb. While \bpel \hspace{0mm} correlation sets allow for multi-party choreographies to be defined, we argue that similar expressivity is attainable with session types by extending them to support multi-party sessions. This challenge for future work has been addressed outside the context of \bpel \hspace{0mm} in \cite{HYC08,Bruni2008}. In \cite{Viroli200774} a calculus based on process algebra and enhanced with a system for correlating related operations is presented. This calculus was shown to be able to reach a certain degree of \bpel -like expressiveness.

%


Blite \cite{Lapadula2012189} is a formalized subset of \bpel \hspace{0mm} with operational semantics that take into account correlation sets. The authors have also written an associated translator that converts Blite code into executable \bpel \hspace{0mm}. Blite uses a \wsdl -typing system, but it does not feature control links. 

	


\vspace{-5mm}

\section{Conclusion}
\vspace{-3mm}

In order to provide a basis for formal reasoning and  verification of service orchestrations,  we have adapted and formalized a subset of the widely adopted orchestration language \bpel. The resulting formalism, that we call \seb \hspace{0mm} for Sessionized \bpel, supports sessions as first class citizens of the language. The separation of the proposed operational semantics into two steps has allowed a relatively concise semantics to be provided when  compared  to previous approaches. Furthermore our semantics take into account the effect of \bpel \hspace{0mm} control links, which are an essential and often neglected part of the language.

In the sequel, we plan to use \emph{session types} as a means of prescribing the correct structure of an interaction between two partner services during the fulfillment of a service. A typed \seb \hspace{0mm} service will declare the session types that it can provide to prospective partners, while also declaring its required session types.  A verification step will be needed to see  whether or not a service is well-typed, hence answering the question of whether or not the service respects its required and provided types. 

We will also investigate and show how the interaction safety property, that we defined on the basis of the semantics of  \seb \hspace{-0mm}, can be guaranteed in configurations of compatible and well typed services. The formal approach taken with \seb \hspace{0mm} as presented in this paper also opens up the possibility of defining and proving other properties of Web service interactions, such as controllability and progress properties. The study of all these verification aspects is left to future work.

\vspace{-4mm}

%
%

\bibliographystyle{eptcs}
\bibliography{biblio}

\end{document}